\documentclass{vldb}
\usepackage{graphicx}
\usepackage{subcaption}
\usepackage{balance}  
\usepackage{times}
\usepackage{latexsym}
\usepackage[dvipsnames]{xcolor}
\usepackage{colortbl}
\usepackage{url}
\usepackage{topcapt}
\usepackage{booktabs}
\usepackage{floatrow}
\usepackage[ruled,vlined, linesnumbered]{algorithm2e}
\usepackage[noend]{algpseudocode}
\usepackage{amsmath}
\usepackage{amssymb}
\usepackage{multirow}
\usepackage{enumitem}
\usepackage{framed}
\usepackage{inconsolata}
\usepackage{wasysym}
\usepackage[makeroom]{cancel}
\usepackage{siunitx}

\newtheorem{definition}{Definition}[section]
\newcounter{exp}
\newtheorem{example}[exp]{Example}
\newtheorem{theorem}[definition]{Theorem}

\newcounter{prob}
\newtheorem{problem}[prob]{Problem}

\usepackage{relsize}
\usepackage[sort]{cite}

\newcommand{\system}{explain{\smaller[2]3D}\xspace}
\newcommand{\System}{Explain{\smaller[2]3D}\xspace}
\newcommand{\exactcover}{\textsc{ExactCover}\xspace}
\newcommand{\probname}{\textsc{exp-3d}\xspace}
\newcommand{\umassdb}{$D_{\text{\textit{UMass}}}$\xspace}
\newcommand{\ncesdb}{$D_{\text{\textit{NCES}}}$\xspace}
\newcommand{\osudb}{$D_{\text{\textit{OSU}}}$\xspace}
\newcommand{\viewa}{$D_{\text{\textit{IMDb1}}}$\xspace}
\newcommand{\viewb}{$D_{\text{\textit{IMDb2}}}$\xspace}

\usepackage{tikz}
\newcommand{\arrow}[1]{
\parbox{#1}{\tikz{\draw[->](0,0)--(#1,0);}}
}
\newcommand{\smallarrow}{\arrow{2mm}}

\newcommand{\threshold}{\textsc{Threshold}\xspace}
\newcommand{\greedy}{\textsc{Greedy}\xspace}
\newcommand{\formal}{\textsc{FormalExp}\xspace}
\newcommand{\expdiff}{\textsc{Explain{\relscale{0.8}3D}}\xspace}
\newcommand{\serf}{\textsc{RSwoosh}\xspace}

\newcommand{\noopt}{\textsc{NoOpt}\xspace}
\newcommand{\bh}{\textsc{Batch-1000}\xspace}
\newcommand{\bl}{\textsc{Batch-100}\xspace}

\definecolor{LightCyan}{rgb}{0.88,1,1}
\definecolor{LightGray}{gray}{0.9}

\newcommand*{\belowrulesepcolor}[1]{
  \noalign{
    \kern-\belowrulesep
    \begingroup
      \color{#1}
      \hrule height\belowrulesep
    \endgroup
  }
}
\newcommand*{\aboverulesepcolor}[1]{
  \noalign{
    \begingroup
      \color{#1}
      \hrule height\aboverulesep
    \endgroup
    \kern-\aboverulesep
  }
}

\pdfoutput=1

\begin{document}

\title{Explain{\textbf{\relscale{1.55}{3D}}}: Explaining Disagreements in Disjoint Datasets}

\numberofauthors{2} 
\author{
\alignauthor
Xiaolan Wang\\
       \affaddr{University of Massachusetts, Amherst}\\
       \affaddr{College of Information and Computer Sciences}\\
       \email{xlwang@cs.umass.edu}
\alignauthor
Alexandra Meliou\\
	   \affaddr{University of Massachusetts, Amherst}\\
       \affaddr{College of Information and Computer Sciences}\\
       \email{ameli@cs.umass.edu}
}

\maketitle

\begin{abstract}
\looseness -1
Data plays an important role in applications, analytic processes, and many
aspects of human activity. As data grows in size and complexity, we are met
with an imperative need for tools that promote understanding and explanations
over data-related operations. Data management research on explanations has
focused on the assumption that data resides in a single dataset, under one
common schema. But the reality of today's data is that it is frequently
un-integrated, coming from different sources with different schemas. When
different datasets provide different answers to semantically similar
questions, understanding the reasons for the discrepancies is challenging and
cannot be handled by the existing single-dataset solutions.

In this paper, we propose \system, a framework for explaining the \textbf{d}isagreements
across \textbf{d}isjoint \textbf{d}atasets (3D). \System focuses on identifying the reasons for
the differences in the results of two semantically similar queries operating
on two datasets with potentially different schemas. Our framework leverages
the queries to perform a semantic mapping across the relevant parts of their
provenance; discrepancies in this mapping point to causes of the queries'
differences. Exploiting the queries gives \system an edge over traditional
schema matching and record linkage techniques, which are query-agnostic. Our
work makes the following contributions:
(1)~We formalize the problem of deriving optimal explanations for the
differences of the results of semantically similar queries over disjoint
datasets. Our optimization problem considers two types of explanations,
provenance-based and value-based, defined over an evidence mapping, which
makes our solution interpretable.
(2)~We design a 3-stage framework for solving the optimal explanation problem.
(3)~We develop a smart-partitioning optimizer that improves the efficiency of
the framework by orders of magnitude.
(4)~We experiment with real-world and synthetic data to demonstrate that
\system can derive precise explanations efficiently, and is superior to
alternative methods based on integration techniques and single-dataset
explanation frameworks.

\end{abstract}

\section{Introduction}
\label{s:intro}

Data drives modern applications, analytic processes, and business decisions,
heavily influencing many aspects of human activity---from product
recommendations and friend connections, to self-driving car decisions and
election campaign strategies. Understanding data and the results of processes
that operate on data becomes critical in promoting trust in data-driven
decisions and in facilitating debugging and repair of errors~\cite{wang2018explaining}. Even within the
relatively simple setting of relational data and queries, the explosive data
sizes, source heterogeneity, and issues of poor data quality make providing
explanations a challenging problem.

Existing data management solutions that aim to provide explanations for query
results~\cite{roy2014formal, Wu13, Roy2015} have an important limitation: They focus
on a single dataset, where data conforms to a single common schema. However,
modern data rarely conforms to this integrated ideal. More often than not,
datasets evolve separately, under different schemas, and even datasets from
trustworthy sources frequently end up diverging, both in format and content,
causing headaches to downstream applications and users. 
For example, Open Data~\cite{miller2018open}, released by governments and
organizations, is typically of high quality, publicly available, and freely
used and distributed. Such datasets may be related and overlapping, but their
separate production and evolution lead to disagreements that can cause
confusion to users and incorrect analyses.

\begin{example}[academic data disagreement]\label{ex:academic}

We collect two publicly-available academic datasets: the UMass-Amherst dataset
on undergraduate
programs\footnote{{\small\url{https://www.umass.edu/gateway/academics/undergraduate}}},
and the National Center for Education Statistics (NCES)
dataset\footnote{{\small \url{https://nces.ed.gov}: A open dataset presented in simplified schema.}}. Both data sources are reputable
and contain high-quality information. Nevertheless, querying both datasets for
the number of undergraduate degree programs at UMass Amherst yields vastly
different answers.

\smallskip
\noindent
\resizebox{\columnwidth}{!}{
\begin{tabular}{p{9mm} l l}
\toprule
 & \textbf{\normalsize UMass-Amherst data (\umassdb)} &
\textbf{\normalsize NCES data (\ncesdb)}\\
\midrule
\textbf{Schema:}
& Major(Major, Degree, School)
& School(\underline{ID}, Univ\_name, City, Url) \\
& & Stats(\underline{ID}, Program, bach\_degr)\\
\textbf{Query:} 
& $Q_1:$ \texttt{SELECT COUNT(Major)}
& $Q_2:$\texttt{SELECT SUM(bach\_degr)}\\
& \texttt{FROM Major;} 
& \texttt{FROM School, Stats}\\
& & \texttt{WHERE Name = `UMass-Amherst' }\\
& & \texttt{AND School.ID=Stats.ID;}\\
\textbf{Answer:} & 113 & 90\\
\bottomrule
\end{tabular}}

\smallskip
Existing explanation solutions can only be applied with respect to one of
these datasets at a time, by asking questions such as ``Why is the result of
$Q_1$ (resp.\ $Q_2$) high (resp.\ low)?'' But these would not provide
meaningful explanations in this case, as each tuple contributes the same to
the aggregate of $Q_1$, and prioritizing tuples with low \texttt{bach\_degr}
in the provenance of $Q_2$ would be arbitrary, not grounded on the actual
differences with $Q_1$.
\end{example}

Example~\ref{ex:academic} illustrates the predicament of dealing with
disagreements in disjoint datasets and how single-dataset explanation
frameworks fall short. Attempting to use data cleaning~\cite{bohannon2005cost,
kolahi2009approximating, prokoshyna2015combining} and data
fusion~\cite{Bleiholder09, dongnaumann09} techniques towards this problem
meets similar challenges. These techniques attempt to reconcile the datasets,
but are agnostic to the queries of interest, which may very well be
contributors to the discrepancy. Ultimately, our goal is not to reconcile the
differences between two datasets and consolidate them, but rather to explain
the reasons of disagreement between two queries on those datasets, whose
results are expected to be the same.

\begin{figure*}[t]
\small
\centering     
\begin{subfigure}[b]{0.24\textwidth}
		\centering
		\begin{tabular}{lc}
		\multicolumn{2}{l}{\textbf{SQL query $Q_1$:}} \\
		\multicolumn{2}{l}{\texttt{SELECT COUNT(program)}}\\
		\multicolumn{2}{l}{\texttt{FROM $D_1$;}} \\[1ex]
		\multicolumn{2}{l}{\textbf{Dataset $D_1$:}}\\ \toprule
		\textbf{Program}  & \textbf{Degree}\\ \midrule
		Accounting & B.S. \\		
		CS  & B.A. \\
		CS  & B.S. \\
		ECE & B.S. \\
		EE & B.S. \\
		Management & B.A.\\
		Design & B.A.\\\bottomrule
		\end{tabular}
		\vspace{-1mm}
\caption{$Q_1(D_1) = 7$}
\end{subfigure}
\begin{subfigure}[b]{0.24\textwidth}
		\centering
		\begin{tabular}{cl}
		\multicolumn{2}{l}{\textbf{SQL query $Q_2$:}} \\
		\multicolumn{2}{l}{\texttt{SELECT COUNT(Major)}}\\
		\multicolumn{2}{l}{\texttt{FROM $D_2$  WHERE Univ=`A';}} \\[1ex]
		\multicolumn{2}{l}{\textbf{Dataset $D_2$:}}\\ \toprule
		\textbf{Univ} & \textbf{Major} \\ \midrule
		A & Accounting\\		
		A & CSE \\
		A & ECE \\
		A & EE \\
		A & Management \\
		A & Design \\
		B & Art \\\bottomrule
		\end{tabular}
		\vspace{-1mm}
\caption{$Q_2(D_2) = 6$}
\end{subfigure}
\begin{subfigure}[b]{0.24\textwidth}
		\centering
		\begin{tabular}{lc}
		\multicolumn{2}{l}{\textbf{SQL query $Q_3$:}} \\
		\multicolumn{2}{l}{\texttt{SELECT SUM(Num\_bach)}}\\
		\multicolumn{2}{l}{\texttt{FROM $D_3$;}} \\[1ex]
		\multicolumn{2}{l}{\textbf{Dataset $D_3$:}}\\ \toprule
		\textbf{College}  & \textbf{Num\_bach}\\ \midrule
		Business & 2 \\		
		Engineering  & 2 \\
		Computer Science & 1 \\\bottomrule
		\multicolumn{2}{l}{}\\
		\multicolumn{2}{l}{}\\
		\multicolumn{2}{l}{}\\
		\multicolumn{2}{l}{}\\
		\end{tabular}
		\vspace{-1mm}
\caption{$Q_3(D_3) = 5$}
\end{subfigure}
\begin{subfigure}[b]{0.24\textwidth}
		\centering
		\begin{tabular}{l c}
		\multicolumn{2}{l}{\textbf{SQL query $Q_4$:}} \\
		\multicolumn{2}{l}{\texttt{SELECT SUM(Num\_major)}}\\
		\multicolumn{2}{l}{\texttt{FROM $D_4$;}} \\[1ex]
		\multicolumn{2}{l}{\textbf{Dataset $D_4$:}}\\ \toprule
		\textbf{Campus}  & \textbf{Num\_major} \\ \midrule
		South campus & 1\\
		North campus & 2\\
		East campus & 1\\\bottomrule
		\multicolumn{2}{l}{}\\
		\multicolumn{2}{l}{}\\
		\multicolumn{2}{l}{}\\
		\multicolumn{2}{l}{}\\
		\end{tabular}
		\vspace{-1mm}
\caption{$Q_4(D_4) = 4$}
\end{subfigure}
\normalsize
\caption{Four queries, operating on disjoint datasets, for answering the question: \textit{How many undergraduate degree programs are provided by University A?}
However, all queries yield different answers: $Q_1(D_1) = 7, Q_2(D_2) = 6, Q_3(D_3) = 5,$ and $Q_4(D_4) = 4$. 
}
\vspace{-2mm}
\label{fig:comp}
\end{figure*}

In this paper, we introduce \system,\footnote{Pronounced ``explained''} a
framework for deriving interpretable explanations for the disagreement in the results of
two semantically similar queries.\footnote{In the context of our work,
semantic similarity is subjectively determined by human raters, and assumed as
part of the input. This is analogous to the standards of semantic similarity
in the natural language processing literature~\cite{dolan2005automatically}.}
\System leverages the
queries in coordination with existing schema matching and entity resolution
methods to derive a semantic mapping across the relevant parts of the queries'
provenance. It processes this initial mapping to find optimal provenance-based
(mismatched tuples) and value-based (mismatched values) explanations and
summarizes these explanations to increase understandability. For the
disagreement in Example~\ref{ex:academic}, \system finds that (1)~several
tuples in \umassdb (such as majors ``Equine Management'' and ``Turfgrass
Management'') do not correspond to tuples in \ncesdb, and (2)~there is a
mismatch of contributions for some tuples---for example, ``Computer Science''
is counted twice in $Q_1$ for the distinct B.S. and B.A. degrees, but
``Computer Science'' has \texttt{bach\_degr}=1 in \ncesdb. \System further
analyzes the common properties of the derived explanations to summarize them
as:
(1)~There is a large portion of mismatches for majors with
\texttt{Degree}=``Associate degree'' in \umassdb;
(2)~There are majors with multiple degree types in \umassdb, counted multiple
times by $Q_1$, for which \texttt{bach\_degr}=1 in \ncesdb.

\System addresses the following challenges:

\begin{description}[leftmargin=0mm, labelindent=0mm,  topsep=0mm, itemsep=0mm]
\item[Different schemas.] Data sources often adopt different schemas and may
thus store their data with different granularities. For example, in
Example~\ref{ex:academic}, \umassdb lists each degree program as an individual
tuple whereas \ncesdb stores an aggregate of the degrees in each program in
the attribute \texttt{bach\_degr}. Such differences significantly increase the
difficulty in determining the mapping relationship between tuples in different
datasets.

\item[Missing data mapping.] Data mapping or tuple mapping is essential in
deriving the explanations. However, existing record linkage and entity
resolution techniques~\cite{benjelloun2009swoosh, kopcke2010frameworks}
typically target mapping entities within the same dataset or datasets with
highly similar schemas. In contrast, in our setting, we can leverage the
queries to provide us both with the relevant provenance, and clues of the
matching attributes.

\item[Distinct queries.] Two queries meant to retrieve the same information
across two datasets with different schemas are bound to be structurally
different. Differences in the queries are confounded with differences in
the data and schemas, obscuring the causes of discrepancies and making
deriving explanations more challenging.
\end{description}

We make the following contributions.

\begin{itemize}[leftmargin=0mm, itemindent=3mm,  topsep=0mm, itemsep=-0.5mm]
    \item We introduce the necessary modeling abstractions and formalize the
    problem of deriving optimal explanations for the disagreements between the
    results of two semantically similar queries over two disjoint datasets. We
    identify explanations as one of two types: provenance-based (indicating
    mismatched tuples between the two datasets) and value-based (indicating
    incorrect values in particular tuples). These explanations are defined
    over an \emph{evidence mapping}, which is an explanation of the
    explanations themselves, making our method interpretable. (Section~\ref{sec:prob})
    
    \item We introduce \system as a 3-stage framework for solving our optimal
    explanation problem. The first stage leverages the queries and standard
    schema matching and record linkage methods to derive an initial mapping
    between the relevant provenance data. The second stage, which is the core
    of our approach, models the optimization problem as a mixed integer linear
    program (MILP) and produces a refined \emph{evidence mapping}. This
    mapping, informed by the queries and the datasets, pinpoints the
    discrepancies between the two datasets. The third stage relies on standard
    methods to analyze the common properties of the discrepancies and
    summarize the explanations. (Section~\ref{sec:sol})
    
    \item We propose a smart-partitioning optimizer that breaks the
    optimization problem of \system's second stage into smaller components,
    which can be solved separately, increasing the efficiency and scalability
    of our framework. (Section~\ref{sec:opt})
    
    \item We perform extensive experimental evaluation of \system using
    real-world and synthetic data, comparing it with a state-of-the-art
    single-dataset explanation framework, state-of-the art entity resolution
    approaches, and multiple baselines. Our evaluation shows that \system is
    superior in explanation accuracy compared to the alternatives, and the
    smart-partitioning optimizer is robust to multiple parameter settings and
    increases efficiency by orders of magnitude with little to no loss of
    accuracy. (Section~\ref{sec:experiments})
    
\end{itemize}

\section{Explanations for disjoint data}
\label{sec:prob}

In this section, we use a running example inspired by
Example~\ref{ex:academic} to introduce our concepts and abstractions for
modeling explanations for disagreements in disjoint datasets.

\begin{example} \label{ex:4q}

Figure~\ref{fig:comp} displays four semantically similar queries that answer
the question ``How many undergraduate programs are provided by University A?''
The queries compute the same thing semantically, but they operate on different
datasets, with different schemas: $D_1$ lists the undergraduate programs at
University A and $Q_1$ counts them; $D_2$ lists the majors at multiple
universities and $Q_2$ selects the ones from University A and counts them;
$D_3$ lists the number of bachelor degrees per college at University A and
$Q_3$ sums them; $D_4$ lists the number of majors per campus at University A
and $Q_4$ sums them. While all four queries are correct semantically, they
ultimately yield different results.

Manually, one can easily contrast $Q_1$ and $Q_2$. The Program and Major
attributes are a direct match, and each program in $D_1$ corresponds to a
major in $D_2$ and vice versa, through a one-to-one mapping: `Accounting' to
`Accounting', `CS' to `CSE', `ECE' to `ECE', etc. This reveals that computer
science is counted twice in $Q_1$, for the B.S.\ and B.A.\ degrees, but only
once in $Q_2$, which explains the difference in their results. Moreover, the
mapping of tuples between the two datasets is an interpretable explanation
(evidence) of the explanation itself.

The correspondence between $Q_1$ and $Q_3$ is a little less straightforward,
because the data is stored at different granularities (list of programs vs
aggregates per college). However, the queries are still comparable. The
Program attribute semantically maps to the College attribute in a containment relationship:
each program typically corresponds to a college---Accounting and Management
are part of the Business School, ECE and EE are part of the College of
Engineering, and CS is part of the College of Computer Science. This mapping
reveals that (1)~CS is counted twice in $Q_1$, for the B.S.\ and B.A.\
degrees, but $D_3$ only lists one bachelor degree in the CS College, and
(2)~the Design program is missing from $D_3$.

While we can reason about the differences of $Q_1$, $Q_2$, and $Q_3$, we
cannot compare them with $Q_4$ because the Campus attribute does not
meaningfully correspond in a direct or containment relationship with the other
datasets.

\end{example}

This example highlights several concepts:
(1)~attribute matches and their implications to
(2)~comparability of queries,
(3)~explanations as mismatched tuples or mismatched values, and 
(4)~evidence mappings that support the derived explanations.
We proceed to formalize these concepts and define the problem of deriving
explanations for disagreements in the results of semantically similar queries
over disjoint datasets.

\subsection{Problem input: Queries, data, and matches} \label{sec:prob:input}
In this paper, we focus on queries of the general relational algebra form $Q =
\pi_{o}\sigma_C (X)$, where $X$ can be a single relation or an arbitrary
query, allowing joins, unions, and subqueries; $C$ also allows any operators,
except UDFs. We restrict the projection, $o$, to be either a set of
attributes, $o = \mathcal{A}\subseteq attr(X)$, or one of the five main SQL
aggregate functions (SUM, COUNT, AVERAGE, MAX, MIN), $o = aggr(A_i), A_i\in
attr(X)$. 
Compared to prior work on explanations over a single
database~\cite{chapman2009not, tran2010conquer, Wu13, roy2014formal}, which
mostly focus on flattened queries in select-project-join (SPJ) and
select-project-join-aggregate (SPJA) format, our framework supports a wider
range of queries.

\begin{figure}[t]
\small
\begin{tabular}{p{3.6cm}l} \toprule
\textbf{Notation} & \textbf{Description}\\\midrule
$Q= \Pi_{o}\sigma_c (R)$ & A query over relation $R$ in database $D$. \\
$\mathcal{M}_{attr}(Q_1, Q_2) = (\mathcal{A}_i\phi \mathcal{A}_j)$ & Attribute matches.\\
$\mathcal{M}_{tuple} = \{(t_i, t_j, p), ...\}$ & Tuple matches.\\
$P(A_1, ..., A_k, I)$ or $P$ & The provenance relation of query $Q$.\\
$T$ & Canonical tuples of query $Q$.\\
$t.I$ & The impact of a tuple $t$. \\
$E = (\Delta, \delta|\mathcal{M}_{tuple}^*)$ & Explanations and their evidence. \\
$\Delta = \{t, ...\} \in E$ & Provenance-based explanations.\\
$\delta = \{t.I \mapsto t.I^*\} \in E$ & Value-based explanations.\\
$\mathcal{M}_{tuple}^* \subseteq \mathcal{M}_{tuple}$ & Evidence of a set of explanations.\\
\bottomrule
\end{tabular}
\vspace{-2mm}
\caption{Summary of notations.}
\end{figure}

As Example~\ref{ex:4q} showed, some queries are not comparable ($Q_1$ and
$Q_4$). Reasoning about these cases would require external information, not
derivable by standard matching and linking methods. We cannot derive
explanations for these cases---this appears impossible without external
information---and we focus on comparable queries. As Example~\ref{ex:4q}
highlighted, comparability is determined by semantic mappings that match
attributes of the queries. We formalize these \emph{attribute matches} below.

\begin{definition}[Attribute matches]\label{def:attrmatch} 
Given two queries $Q_1$ over relation $R_1$ and $Q_2$ over relation $R_2$, we
represent the semantic mapping among their attributes as 
\textbf{attribute matches}, denoted with the matching function $\mathcal{M}_{attr}$:
\[
\mathcal{M}_{attr} (Q_1, Q_2) = (\mathcal{A}_i \phi \mathcal{A}_j)
\]
where $\mathcal{A}_i$, $\mathcal{A}_j$ are sets of categorical attributes in $R_1$ and $R_2$, respectively,
and $\phi \in \{\equiv, \sqsubseteq, \sqsupseteq\}$ is the semantic relation between two sets of attributes~\cite{giunchiglia2004s}.
\end{definition}
In our definition of matching attributes, we borrow the notion of the semantic
relation $\phi$ from prior work~\cite{giunchiglia2004s}. A set of attributes
$A_i$ can be \emph{semantically equivalent} to $A_j$ ($\mathcal{A}_i\equiv
\mathcal{A}_j$), corresponding to a one-to-one mapping between instantiations
of $A_i$ and $A_j$, \emph{less general} than $A_j$ ($\mathcal{A}_i\sqsubseteq
\mathcal{A}_j$), corresponding to a many-to-one mapping, or \emph{more
general} than $A_j$ ($\mathcal{A}_i\sqsupseteq \mathcal{A}_j$), corresponding
to one-to-many mapping.
Note that semantic equivalence does not imply a condition on cardinality and
two semantically equivalent sets of attributes may in fact have arbitrary
overlap. For example, the sets $A_i=(address, city, state, zip)$ in $R_1$ and
$A_j=(address)$ in $R_2$ can be semantically equivalent ($A_i\equiv A_j$).
In our running example, $\mathcal{M}_{attr}(Q_1,Q_2) = (\text{program})\equiv
(\text{major})$, and $\mathcal{M}_{attr}(Q_1,Q_3) =
(\text{program}) \sqsubseteq (\text{college})$.
The attribute matches can be derived from standard schema matching
techniques~\cite{giunchiglia2004s, Berlin2002, Aumueller2005, Zhang2013}.
Deriving these matches is not a focus in our work, and we treat them as part
of our input.  

One can consolidate or separate matches over sets of attributes, e.g.,
$(\text{zip, city}) \sqsubseteq (\text{county})$ becomes $(\text{zip})
\sqsubseteq (\text{county})$ and $(\text{city}) \sqsubseteq (\text{county})$.
Our framework applies to both cases. From here on, for ease of exposition, we
will assume that the attribute matches are on a single attribute from each
relation, and we will simply denote them with $\mathcal{M}_{tuple}$ when the
queries are clear from the context.

If there exists at least one attribute match between two queries, we
can derive explanations for their differences (comparable queries); otherwise,
the queries are not comparable ($Q_1$ and $Q_4$).

\begin{definition}[Comparable queries]\label{def:comparable}
Two queries $Q_1$ over relation $R_1$ and $Q_2$ over relation $R_2$ are
comparable if and only if $\mathcal{M}_{attr}(Q_1,Q_2) \neq \emptyset$.
\end{definition}

We focus on comparable queries in this work, and from here on we will assume
that the queries we discuss are comparable. To derive explanations for query
disagreements, we need to analyze the contents of the two datasets and reason
about their correspondence. We do not need to do so for the entire datasets,
but rather for the parts that contribute to the queries (provenance). For
example, in $Q_2$ only the tuples in $D_3$ with Univ=`A' are part of the
provenance. To facilitate exposition, we derive a \emph{provenance relation}.

\begin{definition}[Provenance Relation]\label{def:provenance}
Given a query $Q=\pi_o\sigma_c(R)$ over relation $R(A_1,\dots)$, we derive a \textbf{provenance relation}
$P(A_1, \dots, I)$ as follows:
For each tuple $t\in\sigma_c(R)$, we create a tuple $t'=(t,I)$ in P, where
$t'.I = \Pi_{o'}(t)$, with $o'=1$ if $Q$ is a non-aggregate query, and $o'=o$
otherwise. The impact of a tuple measures its statistical contribution to the
result of query $Q$.

\end{definition}

In our running example, the provenance relation of $Q_1$ has 7 tuples (same as
$D_1$), each with impact 1; the provenance relation of $Q_3$ has 3 tuples (same
as $D_3$), with impacts 2, 2, and 1, same as the corresponding values of the
Num\_bach attribute.

Given two queries, the tuples of their provenance relations can be associated
through mappings such as the ones described in Example~\ref{ex:4q}. We
formalize the tuple mapping below.

\begin{definition}[Tuple mapping]\label{def:tuplematch}
Given relations $R_1$ and $R_2$, the \textbf{tuple mapping} between $R_1$
and $R_2$ is a set of tuple matches:
\[
\mathcal{M}_{tuple} = \{(t_i, t_j, p), ...\}
\]
where $t_i\in R_1$ and $t_j \in R_2$ are two tuples, and $p\in(0, 1]$ is the
probability that tuple $t_i$ and tuple $t_j$ correspond to the same or
associated (with respect to containment) entities.
\end{definition}

In Example~\ref{ex:4q}, a possible tuple mapping between $Q_1$ and $Q_2$ can
be (omitting superfluous attributes for simplicity) $\mathcal{M}_{tuple} =
\{$(Accounting, Accounting, 1.0), (CS, CSE, 0.9), (ECE, ECE, 1.0), (EE, EE,
1.0), (Management, Management, 1.0), (Design, Design, 1.0)$\}$.
Deriving such matches can be done with traditional
record linkage techniques~\cite{davis2005establishing, benjelloun2009swoosh,
whang2010entity, bhattacharya2006latent,domingos2004multi}. We use such
techniques as blackbox components in our framework to derive an initial tuple
mapping. This initial mapping is typically crude, with many possible tuple
matches of varied probabilities, and it needs to be refined into the correct
mapping $\mathcal{M}_{tuple}^*$. This refinement is a core part of our
framework, which we will discuss in Section~\ref{sec:sol}.

\subsection{Problem output: The explanations} \label{sec:prob:exp}

Example~\ref{ex:4q} highlighted the two generic types of explanations we derive:
(1)~provenance-based explanations, indicating mismatched tuples between the
two datasets, and
(2)~value-based explanations, indicating incorrect values or contributions for
particular tuples.
We formalize these explanations below.

\begin{definition}[Explanations]\label{def:exp}
Given two queries $Q_1$ and $Q_2$ and their provenance relations $P_1$ and
$P_2$, the explanations of their differences include two generic types:
\begin{itemize}[leftmargin=3mm, topsep=2mm, itemsep=0mm] \item
\textbf{A provenance-based explanation} is a tuple $t\in P_1$ (resp.\ $t\in P_2$)
such that $t$ does not map to a
$t'\in P_2$ (resp.\ $t'\in P_1$).  We use $\Delta$ to denote a set of provenance-based explanations.
\item \textbf{A value-based explanation} specifies an impact value change, $t.I
\mapsto t.I^*$, for a tuple $t\in P_1\cup P_2$, meaning that $t$ should have
impact $t.I^*$ rather than $t.I$. We use $\delta$ to denote a set of
value-based explanations.
\end{itemize}
\end{definition}

Example~\ref{ex:4q} highlights a provenance-based explanation for the
disagreement of $Q_1$ and $Q_3$ (the Design program is missing from $D_3$),
and a value-based explanation ($D_3$ only lists one bachelor degree in the CS
College, when it should be two). The derived explanations are tightly coupled
with the tuple mapping. In comparing $Q_2$ and $Q_3$, a mapping that matches
CSE with the Computer Science College, will produce different explanations
than a mapping that matches CSE to the College of Engineering. Typically, the
initial mappings ($\mathcal{M}_{tuple}$) derived from standard entity
resolution and linkage techniques are probabilistic, and would assign the two
possible matches for CSE with two distinct probabilities. Our goal is to discover
the right mapping $\mathcal{M}_{tuple}^*$, leading to the correct (optimal)
set of explanations; we call this refined mapping the \emph{evidence mapping}
(or evidence for short). The evidence mapping is a subset of the initial
mapping ($\mathcal{M}_{tuple}^* \subseteq \mathcal{M}_{tuple}$), and needs to
conform to certain properties discussed in Section~\ref{sec:sol}.

The final product of our framework is a set of explanations and their
evidence, reported as $E = (\Delta, \delta |\mathcal{M}_{tuple}^*)$. The
evidence $\mathcal{M}_{tuple}^*$ is an explanation of the explanations
themselves, making our result fully interpretable.

\subsection{Optimal explanations for 3D}\label{sec:prob:prob}

We now define the problem of deriving optimal explanations for disagreements in disjoint data, which we will refer to as \probname.

\begin{problem}[The \probname problem]\label{def:prob}
Given two queries $Q_1$ and $Q_2$ with provenance relations $P_1$ and $P_2$, respectively, 
and a set of initial tuple matches $\mathcal{M}_{tuple}$,
our goal is derive a set of explanations,
$E = (\Delta, \delta |\mathcal{M}_{tuple}^*)$ that \textbf{maximize the probability}:
\[Pr(E|P_1, P_2, \mathcal{M}_{tuple})\]
\end{problem}

More informally, we are looking for the set of explanations and their evidence
mapping that are the most likely, given the queries' provenance and the
initial probabilistic tuple mapping. In our running example, suppose that the
initial mapping for $Q_2$ and $Q_3$ assigns two possible matches for CSE,
Computer Science and Engineering, each with some probability. This indicates
two possible cases for $\mathcal{M}_{tuple}^*$, mapping CSE to Computer
Science in one case and Engineering in the other. The former choice results in
a single provenance-based explanation (the tuple with major=`Design' in $D_2$
does not have a match in $D_3$). The latter choice, results in the same
explanation and, in addition, that the tuple with College=`Computer Science'
in $D_3$ does not have a match in $D_2$, and that the Num\_bach value of the
Engineering tuple in $D_3$ is wrong. Clearly, the former choice is better.
Intuitively, a particular tuple mapping identifies specific discrepancies,
which we map to explanations, and fewer discrepancies are typically preferred.

In Section~\ref{sec:sol:stage1}, we analyze the calculation of the objective
function of Problem~\ref{def:prob}, and reduce it to a simpler scoring
function that is both tractable and theoretically-grounded. In
Section~\ref{sec:sol:stage2}, we describe a framework for deriving the
explanations and evidence mapping through a translation to Mixed Integer
Linear Programs (MILP). Then, in Section~\ref{sec:sol}, we describe a
smart-partitioning optimizer that improves the efficiency of our basic
approach by several orders of magnitude.

\section{Deriving Explanations}
\label{sec:sol}

\looseness-1
In this section, we present \system, a 3-stage framework that solves
Problem~\ref{def:prob}. The first stage (Section~\ref{sec:sol:stage1}) refines
the provenance data into a canonical form that is easier to analyze. With data
in this canonical form, we define essential properties for evidence
mappings and explanations, and use them to simplify the objective function of
Problem~\ref{def:prob}. The second stage (Section~\ref{sec:sol:stage2}), which
is the core of our approach, models the optimization problem as a mixed
integer linear program (MILP) and produces a refined \emph{evidence mapping}
and the corresponding explanations. The third stage
(Section~\ref{sec:sol:stage3}) relies on standard methods to analyze the
common properties of the discrepancies and summarize the explanations.

\subsection[Stage 1: Canonicalization and Simplification]{\resizebox{0.9\columnwidth}{!}{Stage 1: Canonicalization and Simplification}}\label{sec:sol:stage1}

\looseness-1
The provenance relation $P_1$ of $Q_1$ has two tuples for the CS program, one
for the B.S.\ and one for the B.A.\ degree. The degree information is not
relevant to the comparison with $Q_2$, and it is not part of the mapping
between $Q_1$ and $Q_2$. Thus the two CS tuples in $P_1$ are indistinguishable
with respect their role in the disagreement between $Q_1$ and $Q_2$. This
indicates that the provenance relation contains redundancy. We consolidate
redundant tuples and their impact through \emph{canonicalization}.
Canonicalization groups tuples with the same values for the matching
attributes and sums their impacts. Canonicalization does not change the 
provenance relations of queries that require a strict one-to-one mapping (queries with AVG/MAX/MIN aggregation). The canonical relation of $Q_1$ has 6
tuples (instead of 7 in $P_1$), and CS is represented by a single tuple with
impact 2 (Figure~\ref{fig:canon1}).

\begin{definition}[Canonical Relation] 

Given a provenance relation $P$ of a query $Q$, and attribute matches $\mathcal{M}_{attr}$,
the canonical relation $T$ of $P$ is derived with the query:
\[
T = \pi_{\mathcal{A},I}(_{\mathcal{A}}\mathcal{G}_{\text{\textit{SUM}}(I)}(P))
\]
\looseness -1
Where $\mathcal{A}$ is a set of matching attributes that appear in $\mathcal{M}_{attr}$; 
$_{\mathcal{A}}\mathcal{G}_{\text{SUM}(I)}$ is the Group By operation over attributes $\mathcal{A}$ with aggregate function \textit{SUM} on the impact attribute $I$.
\end{definition}

\begin{example}
Figure~\ref{fig:ctuples} shows the canonical relations of $Q_1$ and $Q_2$
based on the attribute matches $\mathcal{M}_{attr} = (\text{program} \equiv
\text{major})$. The canonical relation of $Q_1$ is constructed with the query:\\
\texttt{SELECT} program, COUNT(I) \texttt{AS} I \texttt{FROM} $P_1$ \texttt{GROUP BY} program
\end{example}

Canonicalization simplifies the datasets without losing information necessary
for the reasoning on disagreements. It further allows us to identify and
formalize essential properties for explanations and evidence mapping, which
we analyze next.

\subsubsection*{Explanation properties}

\textbf{Completeness.}
Explanations define refinements on the canonical relations. A provenance-based
explanation indicates the removal of tuples, and a value-based explanation
modifies a tuple's impact. Our goal is to identify a set of explanations that
is \emph{complete}: if one performs all the refinements defined by the
explanations, the queries would return the same result. We evaluate
completeness through the properties of valid mapping and equal impact.
In the following, we
denote $T_1^* = \delta(T_1\setminus \Delta)$ and $T_2^* = \delta(T_2\setminus
\Delta)$ as the refined tuples of the canonical relations.

\smallskip

\noindent \textbf{Mapping validity.}
The attribute matches ($\mathcal{M}_{attr}$) between two queries imply the
cardinality of the tuple matches between the two canonical relations. If two
attributes have an equivalence match, e.g., $\text{program}
\equiv\text{major}$, then the canonical relations should have a one-to-one
mapping of their tuples. Thus, in Figure~\ref{fig:ctuples}, each tuple in
$T_1$ should map to one tuple in $T_2$. If it is a less general match, e.g.,
$\text{program} \sqsubseteq \text{college}$, then the mapping should be
many-to-one (many programs map to one college). We can never have many-to-many
mappings.

Initial tuple mapping, however, typically do not conform to the required
cardinality, as they frequently assign several probabilistic matches for each
tuple. For example, the CSE major in $Q_2$ may be mapped to two colleges in
$Q_3$, Engineering and Computer Science, which violates the many-to-one
cardinality requirement for two relations. Our goal is to produce a refined
mapping $\mathcal{M}_{tuple}^*$ that conforms to the cardinality requirements
of the attribute matches $\mathcal{M}_{tuple}^*$; we call such a mapping
\emph{valid}.

\begin{definition}[Valid Mapping] \label{def:validmapping}
Given attribute matches $\mathcal{M}_{attr}=(\mathcal{A}_i\phi \mathcal{A}_j)$, 
and two sets of refined tuples, $T_1^*$ and $T_2^*$, the mapping $\mathcal{M}_{tuple}^*$ is \textbf{valid} if and only if the following are true:

\begin{itemize}[leftmargin=3mm, topsep=0mm, itemsep=0mm, parsep=1mm]  
    
\item If $\mathcal{A}_i\sqsubseteq \mathcal{A}_j$, then
$\forall t \in T_1^*, |\{t|(t, t', p) \in \mathcal{M}_{tuple}^*\}| \leq 1 $

\item If $\mathcal{A}_i\sqsupseteq \mathcal{A}_j$, then
$\forall t \in T_2^*, |\{t|(t', t, p) \in \mathcal{M}_{tuple}^*\}| \leq 1 $

\item If $\mathcal{A}_i\equiv \mathcal{A}_j$, then both the above conditions hold.

\end{itemize}
\end{definition}

\noindent \textbf{Impact equality.} 
Tuples of the canonical relations and their mapping form a bipartite graph. In
a valid mapping, where the matches can only be one-to-one, one-to-many, or
many-to-one, the graph separates into connected components. Each component
contains the tuples that correspond to each other semantically. When the two
query results agree, the total impact on each side of the bipartite graph is
the same within each connected component. Thus, our goal is to find a set
of explanations, such that the refined canonical relations $T_1^*$ and $T_2^*$
demonstrate such impact equality.

\begin{figure}[t]
\small
\centering     
\begin{subfigure}[b]{0.49\textwidth}
		\begin{minipage}[t]{0.49\textwidth}
		\centering
		\begin{tabular}{cl|c}
		\toprule
		\textbf{rowID} & \textbf{Program}  & \textbf{\textbf{$I$}}\\ \midrule
		$p_1$ & Accounting &  1\\
		$p_2$ & CS  & 2\\	
		$p_3$ & ECE & 1\\	
		$p_4$ & EE & 1 \\
		$p_5$ & Management & 1\\
		$p_6$ & Design	& 1\\
		\bottomrule
		\end{tabular}
		\end{minipage}
\caption{$T_1$: Canonical relation for $Q_1$}\label{fig:canon1}
\end{subfigure}
\begin{subfigure}[b]{0.49\textwidth}
		\begin{minipage}[t]{0.49\textwidth}
		\centering		
		\begin{tabular}{cl|c}
            \toprule
            \textbf{rowID} & \textbf{Major}  & \textbf{$I$} \\
            \midrule
			$m_1$ & Accounting &  1\\
			$m_2$ & CSE  & 1\\	
			$m_3$ & ECE & 1\\	
			$m_4$ & EE & 1 \\
			$m_5$ & Management & 1\\
			$m_6$ & Design	& 1\\
            \bottomrule
        \end{tabular}
        \end{minipage}
\caption{$T_2$: Canonical relation for $Q_2$}
\end{subfigure}
\vspace{-2mm}
\caption{Canonical relations for queries $Q_1$ and $Q_2$ of Figure~\ref{fig:comp}. $I$ denotes
the impact of the tuples.}
\label{fig:ctuples}
\end{figure}

\begin{definition}[Impact equality] \label{def:eqcond}
    \looseness-1
Given canonical relations $T_1^*$ and $T_2^*$, and a bipartite graph $G$
formed by a valid mapping $\mathcal{M}_{tuple}^*$ between $T_1^*$ and $T_2^*$,
the impact equality property is satisfied if and only if for all connected
components $(T_1', T_2')$ of $G$:
\[
\sum_{t\in T_1'}(t.I) = \sum_{t\in T_2'}(t.I)
\]
\end{definition}

\begin{definition}[Complete explanations] \label{def:explained}
A set of explanations $E = (\Delta, \delta|\mathcal{M}_{tuple}^*)$ over
canonical relations $T_1$ and $T_2$ is \textbf{complete} if
$\mathcal{M}_{tuple}^*$ is a valid mapping and $T_1^* = \delta(T_1\setminus
\Delta)$ and $T_2^* = \delta(T_2\setminus \Delta)$ satisfy the impact equality
property.
\end{definition}

\subsubsection*{Explanation problem revisited}

The objective function of Problem~\ref{def:prob} maximizes the probability
$Pr(E|P_1, P_2, \mathcal{M}_{tuple})$. This probability can be equally and
more efficiently computed over the canonical relations, which are a (lossless,
for the purposes of this problem) summary of the provenance relations:
$Pr(E|P_1, P_2, \mathcal{M}_{tuple})=Pr(E|T_1, T_2, \mathcal{M}_{tuple})$.

From Bayesian inference, this is proportional to the product of three probabilities:
\vspace{-3mm}

\begin{align}\label{prob:all}
Pr(E|&T_1, T_2, \mathcal{M}_{tuple}) \nonumber\\
&\propto Pr(T_1, T_2|E) Pr(\mathcal{M}_{tuple} |T_1, T_2, E) Pr(E)
\end{align}
We next consider each of the three probabilities separately.

\smallskip

\noindent
$\mathbf{Pr(T_1, T_2|E).}$  Assuming that tuples are independent, we have:
\begin{equation}\label{prob:tuple1}
Pr(T_1, T_2|E) = \prod_{t\in T_1\cup T_2}Pr(t|E)
\end{equation}
We use $\alpha$ and $\beta$ to denote the a priori probabilities that $t\in
T_1\cap T_2$ and that $t$ has correct impact $t.I$, respectively. Intuitively,
$\alpha,\beta \in (0.5, 1]$, as a tuple is more likely to be covered by both
queries and have correct impact than not.\footnote{For simplicity, we assume
the same $\alpha$ and $\beta$ for all tuples, but our framework can handle
different values across tuples.}
We then compute the probabilities of the different cases of $t$'s inclusion in a set of explanations $E$ as:

\vspace{-3mm}
\begin{eqnarray}\label{prob:tuple2}
&Pr(t|t\notin \Delta, t \notin \delta)& = \alpha \beta;\ \ \ \ Pr(t|t\notin \Delta, t \in \delta) = \alpha(1-\beta);\nonumber\\
&Pr(t|t\in \Delta, t \notin \delta)& = 1 - \alpha; Pr(t|t\in \Delta, t \in \delta) = 0.
\end{eqnarray}
$Pr(T_1, T_2|E)$ is then derived from Equations~\eqref{prob:tuple1}-\eqref{prob:tuple2}.  
Larger $\Delta$ and $\delta$ lead to lower probabilities, thus the computation
prioritizes smaller provenance- and value-based explanations.

\smallskip

\noindent
$\mathbf{Pr(\mathcal{M}_{tuple} |T_1, T_2, E).}$
Assuming independence in tuple matches:
\begin{equation}\label{prob:match1}
Pr(\mathcal{M}_{tuple} |T_1, T_2, E) = \prod_{m\in \mathcal{M}_{tuple}}Pr(m|T_1, T_2, E)
\end{equation}
\vspace{-4mm}

In addition, for a tuple match $m = (t_i, t_j, p)$, the probability that tuples $t_i$ and $t_j$ match is $p$, thus:
\begin{align}\label{prob:match2}
&Pr(m|m \in \mathcal{M}_{tuple}*, t_i, t_j \in T_1\cup T_2) = p; \nonumber \\
&Pr(m|m \notin \mathcal{M}_{tuple}*, t_i, t_j \in T_1\cup T_2) = 1-p; \nonumber  \\
&Pr(m| t_i, t_j \not\in T_1\cup T_2) = 0.
\end{align}
$Pr(\mathcal{M}_{tuple} |T_1, T_2, E)$ is then derived from
Equations~\eqref{prob:match1}-\eqref{prob:match2}.

The probability computation prioritizes tuple matches with higher
probabilities in the evidence mapping.

\smallskip

\noindent
$\mathbf{Pr(E)}.$
In this paper, we simply set the prior probability of a set of explanations $E$, based on
whether it is complete (Definition~\ref{def:explained}). If $E$ is complete,
then $Pr(E)=1$; otherwise, $Pr(E)=0$. These priors force our framework to only
consider explanations that resolve all disagreements.

\smallskip

We can then compute the objective function from Equation~\eqref{prob:all}.
In practice, to improve efficiency we calculate and later optimize the probability
in the logarithmic space:
\begin{align}\label{prob:objective}
&\log(Pr(E|T_1, T_2, \mathcal{M}_{tuple})) \propto \nonumber\\
&\text{\hspace{1cm}}\log(Pr(T_1, T_2|E)) + \log(Pr(\mathcal{M}_{tuple}|T_1, T_2, E)).
\end{align}
\vspace{-4mm}

Through a reduction from the Exact Cover problem\footnote{The Exact Cover problem is one of Karp's 21 NP-complete problems\cite{karp1972}.}, we can prove that Problem~\ref{def:prob} is NP-complete.
\begin{theorem}\label{thm:NPcomplete}
\probname  (Problem~\ref{def:prob}) is NP-complete.
\end{theorem}
\begin{proof}
\looseness -1
We proove that the \probname problem is NP-complete by reducing from the Exact Cover problem. Let $\mathcal{S}$ be a collection of subsets over 
a set of elements $\mathcal{X}$, the Exact Cover problem is a decision problem that
determines whether there exist a subcollection $\mathcal{S}' \subseteq \mathcal{S}$ such
that each element in $\mathcal{X}$ is covered by exactly one subset in $\mathcal{S}'$.

Given an instance of the Exact Cover problem, we construct a instance of the \probname problem as the follows.
For each element $x_i \in \mathcal{X}$, create a tuple $t_i$ with impact $1$ in $T_1$; for each
subset $S_j\in \mathcal{S}$, create a tuple $t_j$ with impact $|S|$ in $T_2$; create a mapping
from $t_i$ to $t_j$ if the corresponding element $x_i$ is covered by the subset $S_j$. For all tuples in $T_1$, assign $\alpha = 0, \beta = 0$; for all tuples in $T_2$, assign $\alpha = 0.5, \beta
 = 0$; for all mappings, assign $p = 0.5$.
 
If the constructed Problem~\ref{def:prob} has the maximum probability $P(E)>0$, then there exist a cover for the Exact Cover problem. This is because a valid set of explanations would cover all tuples in $T_1$ since $\alpha = 0$ and the impacts or the degree for the tuples in $T_1$ are $1$ since $\beta = 0$. Therefore, the corresponding elements $\mathcal{X}$ are completely covered by the selected subsets exactly once.

If there exist a cover in the Exact problem, the maximum probability of the constructed Problem~\ref{def:prob} 
is above 0. When there exist a cover $\mathcal{S}'$ in the Exact problem, we may create a set of explanations as $E = \{\{t_j|S_j \notin \mathcal{S}'\}, \emptyset, \{(t_i, t_j, p)|S_j \notin \mathcal{S}'\}\}$ and the probability of this set of explanations is $P(E)> 0$.

\noindent Therefore, the \probname problem is NP-Complete. 
\end{proof}

\subsection{Stage 2: MILP transformation}\label{sec:sol:stage2}

In this section, we show how stage 2 of \system transforms the \probname
problem into a mixed integer linear program (MILP). This transformation allows
\system to use modern constrained optimization solvers to derive the optimal
explanations. Later, in Section~\ref{sec:opt}, we show how to optimize
computation in this stage, through a smart-partitioning optimizer.

To translate an instance of the \probname problem into a MILP problem, 
we first convert tuples, their tuple matches, and the associated explanations into linear 
constraints; we then express the explanation completeness properties,
using linear constraints; we complete the translation
process by formalizing a linear expression for the probability of the 
explanations.
\subsubsection*{Expressing explanations}
To express the explanations, we first introduce a binary variable for 
each tuple $t_i \in T_1\cup T_2$ and a binary variable for each tuple match $(t_i, t_j, p)$; we then translate
the changes suggested by the explanations into linear constraints.

\smallskip

\noindent \textbf{Tuple:~} Given a tuple $t_i = (t_i.A_1, ..., I)$, there
are two types of explanations that may be associated with this tuple: (1)~a provenance-based explanation ($t_i\in \Delta$); (2)~a value-based
explanation ($t_i \in \delta$). 
We use a binary variable $x_i$ 
to indicate whether tuple $t_i$ is included in an provenance-based explanation;
To express the value-based explanation, we use a integer variable $t.I^*$ for tuple $t$'s refined impact
and a binary variable $y_{i}$ representing whether the tuple's refined impact is the same as its
original impact ($y_{i} = 1$) or not ($y_{i} = 0$). The binary variable $y_{i}$ should satisfy the following constraint.
\begin{eqnarray}
y_{i} &= &(t.I^* = t.I)
\end{eqnarray}
When $x_i = 1$, the tuple $t_i\in \Delta$ is selected as a provenance-based explanation;
when $x_i = 0$, the tuple $t_i$ remains in the canonical relation and
its impact is set to $t.I^*$.

Based on the binary variables and Equation~\eqref{prob:tuple2}, we express the probability of the explanations being associated with tuple $t_i$ as: 
\begin{align*}
\log(Pr(t_i)) = x_i \otimes a + \underline{(1-x_i) \otimes ((1-y_{i})\otimes b + y_{i}\otimes c)}
\end{align*}
In the above expression, $\otimes$ represents regular multiplication; we
prefer to use $\otimes$ to indicate that it is the semi-module multiplication
by scalars; $a = \log(1-\alpha), b = \log(\alpha) + \log(\beta),$ and $c = \log(\alpha) + \log(1-\beta)$ as three constant values.
Note that the above Equation is quadratic due to the underlined expression: $P_i = (1-x_i) \otimes ((1-y_{i})\otimes b + y_{i}\otimes c)$.  We linearize
$P_i$,
with the help of two constant numbers $L$ and $U$ as  follows~\cite{bazaraa2011linear}.
\begin{align}\label{eq:tuple2}
&P_i \geq L\otimes(1-x_i) \nonumber\\
&P_i \leq U\otimes (1-x_i) \nonumber \\
&P_i \geq (1-y_{i})\otimes b + y_{i}\otimes c - U\otimes x_i \nonumber \\
&P_i \leq (1-y_{i})\otimes b + y_{i}\otimes c - L\otimes x_i \nonumber \\
&\log(Pr(t_i)) = x_i \otimes p_1 + P_i
\end{align}
The constant number $L$ (or $H$) cannot be greater than the lower bound (or smaller than the upper bound) of $P_i$.
\smallskip

\noindent \textbf{Tuple Match:~} Given a tuple match $m = (t_i, t_j, p)$, we use 
a binary variable $z_{i,j}$ to express whether it is a true match:
When $z_{i,j} = 1$, we include it in the evidence mapping.
The probability of this match is computed as follows.
\begin{align}\label{eq:mappingexp}
&z_{i,j} \leq (1-x_i);\text{\hspace{1cm}} z_{i,j} \leq (1-x_j) \nonumber \\
&\log(Pr(m)) = z_{i,j} \otimes \log(p) + (1-z_{i,j})\otimes log(1-p)
\end{align}
Where $x_i$ and $x_j$ are the binary variables associated with $t_i$ and $t_j$, respectively.

\subsubsection*{Expressing explanation completeness}
We use the explanation variables to express the mapping validity and impact
equality properties as linear constraints.

\smallskip

\noindent \textbf{Valid Mapping:~}
As required by Definition~\ref{def:validmapping}, the refined 
tuple matches $\mathcal{M}_{tuple}^*$ should follow the valid mapping property,
which essentially restricts the degree
for some of the tuples to be less than or equal to $1$. If $t_i$ is such a tuple, then we add the constraints:
\begin{align}
 \sum_{(t_i, t_j, p) \in M} z_{i,j} \leq 1 \label{eq:validmapping}
\end{align}

\smallskip

\noindent \textbf{Equal Impact:~} 
Valid mappings between the canonical tuples $T_1$ and $T_2$ can never have
many-to-many cardinality. Therefore, in the bipartite graph between $T_1$ and
$T_2$ under a valid mapping, at least one of $T_1$ or $T_2$ is guaranteed to
have only tuples with maximum degree of 1. This observation allows us to
simplify the specification of the connected components in the bipartite graph
and the corresponding impact calculations.  Suppose that all tuples in $T_1$ have maximum degree 1. Then the set of connected components is:
\[
\mathcal{S} = \{(\eta(t_j), t_j, M)|t_j \in T_2\}
\]
where $\eta(t_j)$ is the set of $T_1$ tuples that are adjacent to $t_j \in T_2$.
Consider one connected component $(\eta(t_j), t_j, M) \in \mathcal{S}$, the total impact
of $T_1$ in the component is $I_{l} = \sum_{t_i \in \eta(t_j)}z_{i,j} \otimes t_i.I^*$;
and the total impact of $T_2$ tuples is $I_{r} = t_j.I^*$.
Here, we linearize the quadratic equation $I_l$ using the same method as Equation~\eqref{eq:tuple2}.
\begin{align}\label{eq:multi} 
&I_i \leq U \otimes z_{i,j} \nonumber\\
&I_i \geq L \otimes z_{i,j} \nonumber \\
&I_i \leq t_i.I^* - L \otimes (1-z_{i,j}) \nonumber \\
&I_i \geq t_i.I^* - U \otimes (1-z_{i,j})
\end{align}
Where $I_i = z_{i,j} \otimes t_i.I^*$ is an element in $I_l$; $L$ and $U$ are two constants that cannot be greater than the lower bound (or smaller than the upper bound) of a tuple's impact.

Finally, the equal impact property requires:
\begin{equation} \label{eq:multi2}
\sum_{t_i \in \eta(t_j)} I_i = I_l
\end{equation}

\subsubsection*{Formalizing the objective function}
The \probname problem aims to derive a set of complete explanations such that the probability of the explanations is maximized. 
The MILP formulation creates variables for all provenance-based (tuples) and all value-based (impact) explanations.
Our objective function can be formulated as
a linear expression over the explanation variables in a fashion similar to the constraints of the explanation properties: 
\begin{eqnarray}
\log(Pr(E|\mathcal{T}, \mathcal{M})) = \sum_{t\in \mathcal{T}} \log(Pr(t)) + \sum_{m\in \mathcal{M}}\log(Pr(m))
\end{eqnarray}
Where $\mathcal{T} = T_1\cup T_2, \mathcal{M} = \mathcal{M}_{tuple}$; $\log(Pr(t))$ and $\log(Pr(m))$ are formulated by Equation~\eqref{eq:tuple2} and Equation~\eqref{eq:mappingexp} respectively.

\subsubsection*{The algorithm}

\begin{algorithm}[t]
\caption{The basic solution}
\label{alg:basic}
\small
\SetKwInOut{Input}{Input}
\SetKwInOut{Output}{Output}
\Input{Two sets of canonical tuples ($T_1, T_2$) and acquired tuple matches ($\mathcal{M}_{tuple}$)} 
\Output{A set of explanations}
\smallskip
$milp\_vars, milp\_cond, prob\_expr \leftarrow \emptyset$\;
\smallskip
\ForEach{tuple $t$ in $T_1\cup T_2$}{
   $milp\_vars \leftarrow milp\_vars\ \cup $  DefineTupleVariables($t$)\; \label{alg:tvariables}
   $milp\_cond \leftarrow milp\_cond\  \cup $ TupleImpactCondition($t$)\; \label{alg:tuplecons}
   $prob\_expr \leftarrow prob\_expr\ \cup $ TupleProbability($t$)\; \label{alg:tupleprob}
}
\smallskip
\ForEach{mapping $m$ in $\mathcal{M}$} {
$milp\_vars \leftarrow milp\_vars\ \cup $  DefineMappingVariables($m$)\; \label{alg:mvariables}
   $prob\_expr \leftarrow prob\_expr\ \cup $ MappingProbability($m$)\; \label{alg:mappingprob}
}
\smallskip
$milp\_cond \leftarrow milp\_cond\  \cup $ FormConditions($milp\_vars$)\; \label{alg:conditions}
$milp \leftarrow$ FormMILP($milp\_cond, prob\_expr$)\; \label{alg:finish}
$solved\_vars \leftarrow$ SolveMILP($milp$)\; \label{alg:solve}
$E \leftarrow $ DecodeVariables($solved\_vars$)\; \label{alg:deriveexp}
\smallskip
\Return $E$\;
\end{algorithm}

Algorithm~\ref{alg:basic} provides the pseudocode implementing the MILP transformation described in this section.
The algorithm first iterates over all tuples in the input to define variables, construct constraints, and
express the tuple probabilities in Lines~\ref{alg:tvariables}-\ref{alg:tupleprob}. The algorithm then iterates over all the tuple matches 
and formalizes the probability expression in Line~\ref{alg:mappingprob} according to Equation~\eqref{eq:mappingexp}.
Next, the algorithm constructs constraints for the completeness requirement, as in Equations~\eqref{eq:validmapping}-\eqref{eq:multi2},
by a \textit{FormConditions} function (Line~\ref{alg:conditions}). With the variables and constraints, the algorithm completes the MILP problem formulation and calls a MILP solver to get a solution (Line~\ref{alg:finish}-\ref{alg:solve}). 
We derive the final explanations from the MILP solution by including an
explanation or evidence (tuple match) if the solve value of the corresponding
binary variable is 1 (Line~\ref{alg:deriveexp}).

\subsection{Stage 3: Summarization}
\label{sec:sol:stage3}

The product of stage 2 of \system is a set of explanations and their evidence
mapping. But if the discrepancies between two datasets are extensive, the
derived explanations could involve a large number of tuples and values.
Reviewing such explanations can be tedious. Stage 3 of our framework is tasked
with summarizing and abstracting the explanations to reduce their size and
increase their understandability.
As in Example~\ref{ex:academic}, 
we may summarize the common patterns of the derived
explanations as \textit{\texttt{Degree}=``Associate degree'' in \umassdb},
which is easier to understand than presenting the explanations individually.

Different summarization methods are possible. \System marks tuples associated
with explanations as a ``target'' and then uses existing techniques, such as
Data Auditor~\cite{GolabKKS10} and Data X-Ray~\cite{wang2015data} to identify
common patterns for the target tuples. Alternatively, ``target'' tuples could
be treated as examples by QBE (Query-By-Example)
techniques~\cite{shen2014discovering, psallidas2015s4, deutch2016qplain,
fariha2018squid}, which can then generate SQL queries that precisely describe
them.
Developing novel summarization methods is not a focus of our work in this
paper, and thus stage 3 relies on existing tools. Detailed stage 2
explanations are still available through \system, for users who prefer to
peruse the more precise and detailed causes of the disagreement.

\section{Partitioning optimization}\label{sec:opt}
A critical problem with stage 2 of the \system framework is that it does not
scale for problems with a large number of tuples and tuple matches. The
problem is that the generated MILP grows to sizes that can stump even
state-of-art solvers. To improve the efficiency of the basic algorithm, we can
split the bipartite graph $G = (T_1, T_2, \mathcal{M}_{tuple})$ into its
maximal connected components and solve the problem in each component
separately. This method requires linear time, $O(|T_1| + |T_2| +
\mathcal{M}_{tuple})$, to derive the connected components and it does not
sacrifice the accuracy. However, it fails to achieve any efficiency or
scalability guarantees, as in the worst case, $G$ may be connected.

Inspired by the connected components approach, we propose a method to divide
the original problem into a collection of sub-problems with bounded sizes such
that each sub-problem is guaranteed to be small enough to solve. Our
partitioning method is based on the Graph Partitioning Problem
(GPP)~\cite{kernighan1970efficient, karypis1999chameleon,
phillips2005scalable, bulucc2016recent}, which aims to minimize the total
weight of the edge cuts\footnote{Edge cuts refer to edges across partitions.}.

\begin{problem}[The Graph Partitioning Problem] \label{prob:partition}
Given a number $k\in \mathbb{N}_{>1}$, a bipartite graph $G = (T_1, T_2, \mathcal{M}_{tuple})$ 
formed by tuples and their matches, and an upper bound $L_{max}$ for the maximum partition size, we seek a partition $\Pi$ of $T_1\cup T_2$ with disjoint collections 
of tuples $\Pi = \{(T_{1,1}, T_{2,1}),..., (T_{1,k}, T_{2,k})\}$ such that:
\vspace{2mm}
\begin{itemize}[leftmargin=5mm, topsep=0mm, itemsep=0mm] 
\item $T_{1,1} \cup ... \cup T_{1,k} = T_1 $   and  $ T_{2,1} \cup ... \cup T_{2,k} = T_2;$
\item $|T_{1,i}| + |T_{2,j}| \leq L_{max};$
\item $\text{EdgeCutSum}(\Pi) = \sum_{(t_i, t_j) \in E} w(t_i, t_j)$ is minimized.
\end{itemize}
Where $E = \{(t_i, t_j), ...\}$ denotes the set of edges across partitions; $w(t_i, t_j)$ denotes
the weight of edge $(t_i, t_j)$; $|T_{1,i}| + |T_{2,j}| \leq L_{max}$ is the balancing constraint over the maximum
size of one partition.
\end{problem}

\looseness -1
In our setting, 
a na\"ive way to assign the edge weights is by using the tuple matches' probabilities:
$w(t_i, t_j) = p.$
However, this setting is ill-suited for our problem:
According to our objective function (Problem~\ref{def:prob}),
cutting a high probability tuple match
tends to hurt our objective value much more than cutting
multiple lower probability tuple matches with equal or even higher total probabilities.
For example, let us assume that we cut a tuple match, with $0.9$
probability, that is part of the optimal explanation ($\mathcal{M}_{tuple}^*$).
The objective value, $Pr(E)$, would drop by $9$ times\footnote{This is based on the assumption that the probabilities of other tuples and tuples matches are not impacted.} as the probability
of this tuple match, $Pr(m|m\in \mathcal{M}_{tuple}^*)$, would change from 0.9 to 0.1. This objective value loss is significantly higher
than cutting two tuple matches with lower individual ($0.6$ each) but higher total probabilities ($1.2$ in total).
The latter case would only lead to a objective value drop by $2.25$ times.
Based on this observation, we prioritize cutting tuple matches with lower probabilities and avoid cutting
tuple matches with high probabilities.
We achieve this by adjusting the edge weight assignments as below:
\[
w(t_i, t_j) = 
\begin{cases}
p\cdot R, &if\ p \geq \theta_h;\\
p/R, & if\ p \leq \theta_l; \\
p,  &otherwise.
\end{cases}
\]
\looseness -1
Where $R \in (1, \infty)$ is a constant for rewarding (or penalizing) high probability (or low probability)
tuple matches and $0 \leq \theta_l < \theta_h \leq 1$ are two thresholds specifying
low and high probability tuple matches. In this paper, 
we set $\theta_l = 0.1, \theta_h = 0.9, R = 100$. 

Existing graph partitioners, e.g., METIS~\cite{phillips2005scalable} and
hMETIS~\cite{karypis1999chameleon}, can be used directly to derive the
sub-problems, but they are not efficient when $R$ is large.
To further optimize partitioning efficiency, we employ a
\textit{pre-partitioning step} that combines tuples connected by high
probability tuple matches. This pre-partitioning step can also be considered
as an extra coarsening level on top of the \textit{multilevel graph
partitioning algorithms}~\cite{karypis1995multilevel, karypis1998fast}.
Empirically, this step achieves $200\times$ partition time speedup over graphs
with $10K$ tuples without compromising optimality.

\begin{algorithm}[t]
\caption{The pre-partitioning algorithm}
\label{alg:prepart}
\small
\SetKwInOut{Input}{Input}
\SetKwInOut{Output}{Output}
\Input{A bipartite graph $G = (T_1, T_2, \mathcal{M}_{tuple})$ and \\thresholds $\theta_l, \theta_h, R$} 
\Output{A simplified graph $G_{c} = (C_1, C_2, \mathcal{M}_{c})$}
\smallskip
$C_1, C_2, \mathcal{M}_{c} \leftarrow \emptyset$\;
\smallskip
\ForEach{tuple $t$ in $T_1\cup T_2$}{ \label{alg:part:start}
   \If{\textit{t.isVisited}} {
   		continue }
   $(T_1', T_2') \leftarrow $ FindHighProbTuplesDFS($t, G, \theta_h$)\; \label{alg:part:dfs}
   $(C_1', C_2') \leftarrow$ MergeTuples($T_1'$, $T_2'$)\; \label{alg:part:comb1}
   $(C_1, C_2)\leftarrow$ UpdateMergedTuples($C_1', C_2'$)\;\label{alg:part:comb2}
} \label{alg:part:end}
\smallskip
\ForEach{mapping $(t_i, t_j, p)$ in $\mathcal{M}_{tuple}$} {\label{alg:part:update1}
   $(C_i', C_j') \leftarrow$ FindMergedTuples($C_1, C_2, t_i, t_j$)\;
   $\mathcal{M}_c \leftarrow$ UpdateEdgeWeight($C_i', C_j', p, R$) \label{alg:part:update2}
}
\Return $G_{c} = (C_1, C_2, \mathcal{M}_{c})$\;
\end{algorithm}

Algorithm~\ref{alg:prepart} presents the pseudocode of the pre-partitioning step.
The algorithm iterates over tuples in the bipartite graph 
in arbitrary order and attempts to merge tuples that are connected by 
high probability tuple matches as much as possible (Lines~\ref{alg:part:start}-\ref{alg:part:comb2}). 
It then iterates over the remaining tuple matches and updates the edge weights of
the merged tuples accordingly (Lines~\ref{alg:part:update1}-\ref{alg:part:update2}).
This algorithm has linear time complexity: $O(|T_1| + |T_2| + |\mathcal{M}_{tuple}|)$.

Finally, Algorithm~\ref{alg:smartpart} presents our smart-partitioning method. This algorithm 
first leverages the pre-partitioning algorithm (Algorithm~\ref{alg:prepart}) to generate a much smaller graph (Line~\ref{alg:smart:prepart}); it
then partitions the smaller graph (Line~\ref{alg:smart:part}) with a standard graph partitioner; it finally produces
the final partitioning $\Pi$ according to the tuples' assigned partitions (Lines~\ref{alg:smart:p1}-\ref{alg:smart:p2}). 

\begin{algorithm}[t]
\caption{The smart-partitioning algorithm}
\label{alg:smartpart}
\small
\SetKwInOut{Input}{Input}
\SetKwInOut{Output}{Output}
\Input{A bipartite graph $G = (T_1, T_2, \mathcal{M}_{tuple})$, thresholds $\theta_l, \theta_h$, $R$, the number of partitions $k$, and the maximum partition size $L_{max}$} 
\Output{A partition $\Pi$}
\smallskip
$G_{c} \leftarrow $ PrePartition($G, \theta_l, \theta_h, R$)\; \label{alg:smart:prepart}
$\Pi_c \leftarrow $ GraphPartitioner($G_c, k, L_{max}$)\; \label{alg:smart:part}
$\Pi \leftarrow $ InitializeKEmptyPartitions($k$) \; \label{alg:smart:p1}
\ForEach{$(C_1', C_2')$ in $G_{c}$} {
	$idx \leftarrow \Pi_c(C_1', C_2')$ \;
	$\Pi[idx] \leftarrow $ AddTuples($C_1', C_2'$)\; \label{alg:smart:p2}
}
\Return $\Pi$\;
\end{algorithm}

\section{Experimental Evaluation}
\label{sec:experiments}

In this section, we evaluate the effectiveness and efficiency of \system
using both real-world and synthetic data. In particular, 
we first compare \system with several alternative algorithms over
two categories of real-world data (Section~\ref{exp:real}); then, we
evaluate the performance and benefit of the smart-partitioning
optimization over a series of synthetic datasets with diverse properties (Section~\ref{exp:synthetic}).

\subsection{Experimental setup}
All experiments were performed on $4 \times 2.77$ GHz machines with
32GB RAM running IBM CPLEX~\cite{cplex} as the MILP solver on
MacOS version 10.11.6.

\subsubsection{Datasets, queries, and gold standards}\label{sec:exp:data}
We first describe the real-world data used in 
our evaluation; we describe our synthetic data experiments in Section~\ref{exp:synthetic}.
\begin{description}[leftmargin=2mm, topsep=0mm, itemsep=0mm]
\item[Academic datasets.] We collect three publicly available academic datasets, 
the UMass-Amherst dataset
on undergraduate
programs and the National Center for Education Statistics (NCES)
dataset, described in Example~\ref{ex:academic}, and the
the OSU dataset on undergraduate programs\footnote{\small\url{http://undergrad.osu.edu/majors-and-academics/majors}}.  We create two pairs 
of datasets for comparisons: (1)~\emph{UMass-Amherst vs. NCES}, described in Example~\ref{ex:academic}, and (2)~\emph{OSU vs. NCES}, described in the table below.
We evaluate all alternative algorithms with queries that compute \emph{the number of undergraduate programs at UMass Amherst (or OSU, respectively)} on each pair of data.

\smallskip
\noindent
{\small \it
\begin{tabular}{l l}
\toprule
\textbf{\normalsize OSU data (\osudb)} &
\textbf{\normalsize NCES data (\ncesdb)}\\ 
\midrule
Major(Major, Degree, Campus, & School(\underline{ID}, Univ\_name, City, Url) \\
 School) & Stats(\underline{ID}, Program, bach\_degr)\\\hline
$Q_1:$  & $Q_2:$ \\
\texttt{SELECT COUNT(Major)} & \texttt{SELECT SUM(bach\_degr)}\\
\texttt{FROM Major;}  & \texttt{FROM School, Stats}\\
& \texttt{WHERE Name = `OSU' }\\
& \texttt{AND School.ID=Stats.ID;}\\
\bottomrule
\end{tabular}}

\smallskip

\looseness-1
\noindent \textbf{Gold Standard:} We manually create the gold standard 
for the explanations and the evidence mapping on both pairs of data.
The datasets, queries, and gold standards are publicly 
available\footnote{{\small{\url{https://bitbucket.org/xlwang/explain3d}}}}. 
Figure~\ref{tab:datastats} shows the detailed statistics of the academic datasets.

\item[IMDb Datasets.] We retrieve the IMDb data\footnote{{\small\url{https://datasets.imdbws.com/}}}, and use it to create a pair of disjoint datasets, as two views with different
schemas over the original data.
To simulate the real-world disagreements over disjoint data, we choose a schema design for 
the first view such that a certain portion of data is lost during the data 
migration process.\footnote{In \viewa, a movie is associated with a single country and genre.}
We further introduce $\sim$5\% random errors
to both views with the BART system~\cite{arocena2015messing}. 
We create 10 query templates (listed below), mapped over each view, covering a wide range of
query types, including joins, subqueries, non-aggregates, and 5 different
aggregate functions. We create 10 instantiations of each template, by
selecting a random value for \emph{year}$\in[1970,2003]$ for templates
$Q_1$--$Q_{9}$, and a random value for \emph{genre} in $Q_{10}$, resulting in
a total of 100 different queries.

\noindent
{\small \it
\resizebox{\columnwidth}{!}{
\begin{tabular}{p{.25cm} l}
\toprule
\multicolumn{2}{c}{\textbf{IMDb View 1 (\viewa)}}  \\\midrule
\multicolumn{2}{l}{Movie (\underline{movie\_id}, title, release\_year, genre, country, runtimes, gross,} \\
\multicolumn{2}{l}{~~~~~~~~~~~~budget)}\\
\multicolumn{2}{l}{Actor (\underline{actor\_id}, firstname, lastname, gender, dob)}\\
\multicolumn{2}{l}{Director (\underline{director\_id}, firstname, lastname, gender, dob)}\\
\multicolumn{2}{l}{MovieDirector (movie\_id, director\_id)~MovieActor (movie\_id, actor\_id)}\\\toprule
\multicolumn{2}{c}{\textbf{IMDb View 2 (\viewb)}}  \\\midrule
\multicolumn{2}{l}{Movie (\underline{m\_id}, title, release\_year) MovieInfo (m\_id, info\_type, info)} \\
\multicolumn{2}{l}{Person (p\_id, name, gender, dob) MoviePerson (m\_id, p\_id)} \\\midrule
\multicolumn{2}{c}{\textbf{Query templates}} \\\midrule
$Q_1$ & Return actors who were cast in short movies released in $\langle$year$\rangle$. \\
$Q_2$ & Return movies directed by someone born in $\langle$year$\rangle$. \\
$Q_3$ & Return the number of comedy movies released in $\langle$year$\rangle$. \\
$Q_4$ & Return the number of movies released in the US in $\langle$year$\rangle$. \\
$Q_5$ & Return the total gross value for movies released in $\langle$year$\rangle$. \\
$Q_6$ & Return the maximum gross value for movies released in $\langle$year$\rangle$. \\
$Q_7$ & Return the longest movie released in $\langle$year$\rangle$. \\
$Q_8$ & Return the average gross value for movies released in $\langle$year$\rangle$. \\
$Q_{9}$ & Return the average runtime for movies released in $\langle$year$\rangle$. \\
$Q_{10}$ & Return actresses who have not starred in any $\langle$genre$\rangle$ movies. \\
\hline
\end{tabular}
}\\
}

\smallskip

\looseness -1
\noindent \textbf{Gold Standard:} While creating the two disjoint views,
we keep track of the data lost in the first view and record the random
errors introduced by BART; these are the optimal explanations of the query disagreements. 
The optimal evidence mapping can also be easily acquired through the mapping 
between the views and the original dataset. The detailed statistics of the 
IMDb datasets are shown in Figure~\ref{tab:datastats}.

\end{description}

\subsubsection{Attribute matches and tuple mapping} \label{exp:tuplematches}
\noindent \textbf{Attribute Matches.} The attribute matches ($\mathcal{M}_{attr}$) for the two real-world datasets
are shown in Figure~\ref{fig:attrmatch}.

\begin{figure}[t]
\caption{Dataset statistics. $N$, $|P|$, $|T|$ are the original data
    size, the provenance relation size, and the canonical relation size,
    respectively; the size of the initial tuple mapping is
    $|\mathcal{M}_{tuple}|$; the sizes of the optimal evidence mapping and the
    optimal explanations are $|\mathcal{M}_{tuple}^*|$ and $|E|$,
    respectively. $|E_S|$ is the size of the explanations after summarizing
    them with Data X-Ray~\cite{wang2015data, WangFWDM15}. $N$ for \viewa and \viewb are 3.7M and 6.8M tuples, respectively, for all IMDb queries. In the IMDb datasets,
    we show the average numbers over 10 instantiations of each query; $|P|$, $|T|$ in these datasets are the same, so we report only one.
    }
 \centering
 \resizebox{\columnwidth}{!}{
 \begin{tabular}{c|p{1.65cm}|p{1.65cm}|p{1.45cm}|p{1.45cm}} 
     \toprule
     \belowrulesepcolor{LightGray}
     \rowcolor{LightGray}
 \multicolumn{5}{c}{\large\textbf{Academic datasets}}\\
 \aboverulesepcolor{LightGray}
 \midrule
 &\multicolumn{2}{c|}{\# of undergrad majors} 
 &\multicolumn{2}{c}{\# of undergrad majors} \\
 \hline
 & \multicolumn{1}{c|}{UMass} & \multicolumn{1}{c|}{NCES} & \multicolumn{1}{c|}{OSU} & \multicolumn{1}{c}{NCES} \\ 
 \hline
 $N/|P|/|T|$ & \multicolumn{1}{c|}{~~113/113/95~~} & \multicolumn{1}{c|}{~~239K/81/81~~} & \multicolumn{1}{c|}{~~282/282/206~~~} & \multicolumn{1}{c}{~~239K/153/153
~~}
\\ 
 \hline
 $|\mathcal{M}_{tuple}|$ & \multicolumn{2}{c|}{169} & \multicolumn{2}{c}{607} \\
 \hline
 $|\mathcal{M}_{tuple}^*|$ & \multicolumn{2}{c|}{71} & \multicolumn{2}{c}{140} \\
 \hline
 $|E|\rightarrow |E_S|$ & \multicolumn{2}{c|}{64 $\rightarrow$ 11} & \multicolumn{2}{c}{127 $\rightarrow$ 16} \\
 \bottomrule
 \end{tabular}}
\resizebox{\columnwidth}{!}{
\begin{tabular}{c|c|c|c|c|c}
\toprule
\belowrulesepcolor{LightGray}
\rowcolor{LightGray}
\multicolumn{6}{c}{\large\textbf{IMDb datasets}}\\
\aboverulesepcolor{LightGray}
\midrule
& $Q_1$&$Q_2$&$Q_3$&$Q_4$&$Q_5$\\
\hline
$|P|$ {\tiny (IMDb1/IMDb2)} &  1.3K/4.6K & 2.8K/3.8K & 1.6K/3.1K & 2.7K/6.2K & 8.9K/9.0K\\\hline
$|\mathcal{M}_{tuple}|$ & 0.6M & 0.8M & 51K & 0.3M& 1.1M\\
\hline
$|\mathcal{M}_{tuple}^*|$ & 1271 & 	2768	 & 	1601 & 	2756	  & 4231\\
\hline
$|E| \rightarrow |E_S|$ &  3.4K\smallarrow 33 & 1.1K\smallarrow 23  & 1.5K\smallarrow 28 & 3.4K\smallarrow 38&  5.5K\smallarrow 43 \\ \midrule
& $Q_6$&$Q_7$&$Q_8$&$Q_9$&$Q_{10}$\\\hline
$|P|$ {\tiny (IMDb1/IMDb2)} &  5.8K/5.9K & 10.9K/10.9K & 3.4K/3.5K & 4.8K/4.9K & 11.5K/14.4K\\\hline
$|\mathcal{M}_{tuple}|$  &	0.5M & 	2.2M &	0.2M &	0.4M& 1.3M \\
\hline
$|\mathcal{M}_{tuple}^*|$ & 5353& 6259& 2365& 3147 &  5959\\
\hline
$|E| \rightarrow |E_S|$ & 1.3K\smallarrow 19 & 21.1K\smallarrow 86 & 2.5K\smallarrow 33 & 3.9K\smallarrow 40 & 13.4K\smallarrow 75\\
\bottomrule 
\end{tabular}}
\label{tab:datastats}
\end{figure}

\begin{figure}[t]
    \small
    \begin{tabular}{c  c}
        \toprule
        \belowrulesepcolor{LightGray}
        \rowcolor{LightGray}
    \textbf{UMass vs.\ NCES} & \textbf{OSU vs.\ NCES} \\
    \aboverulesepcolor{LightGray}
    \midrule
    (Major.Major) $\sqsubseteq$ (Stats.Program) & (Major.Major) $\sqsubseteq$ (Stats.Program) \\ 
    \bottomrule
    \toprule
    \belowrulesepcolor{LightGray}
    \rowcolor{LightGray}
    \multicolumn{2}{c}{\textbf{IMDb View 1 vs.\ IMDb View 2}}\\ 
    \aboverulesepcolor{LightGray}
    \midrule
    \multicolumn{2}{l}{(Movie.title, Movie.release\_year) $\equiv$ (Movie.title, Movie.release\_year)} \\
    \multicolumn{2}{l}{(Actor.firstname, Actor.lastname, $\equiv$ (Person.name, Person.gender,} \\
    \multicolumn{1}{l}{~Actor.gender, Actor.dob) }&   \multicolumn{1}{l}{~Person.dob)}\\
     \multicolumn{2}{l}{(Director.firstname, Director.lastname, $\equiv$ (Person.name, Person.gender,} \\
     \multicolumn{1}{l}{~Director.gender, Director.dob) }&   \multicolumn{1}{l}{~Person.dob)}\\
    \bottomrule
    \end{tabular}
    \caption{Attribute matches for the real-world datasets.}
    \vspace{-2mm}
    \label{fig:attrmatch}
\end{figure}

\smallskip

\noindent \textbf{Tuple Mapping.}
In our evaluation, we use a similarity-to-probability method~\cite{whang2013question, firmani2016online}
to collect the initial tuple mapping ($\mathcal{M}_{tuple}$).
This similarity-to-probability method is a two-step process that generates the tuple matches probabilities 
from their similarity values:
(1)~it first divides the tuple matches into $k$ continuous buckets over the similarity values; (2)~in each
bucket, it calculates the probability of tuple matches by the ratio of true matches within
the current bucket. The true matches can be acquired by labeling a subset of data, or
by a known gold standard.

To generate the similarity values, we use token-wise Jaccard similarity for \emph{String attributes}: 

\vspace{-3mm}
{\small
\[
sim(t_i.A, t_j.A) = \frac{|t_i.A \cap t_j.A|}{|t_i.A \cup t_j.A|}
\]
\normalsize}
We use normalized
Euclidean distance on \emph{numeric attributes}: 

\vspace{-3mm}
{\small
\[
sim(t_i.A, t_j.A) = \frac{1}{1 + |t_i.A - t_j.A|^2}
\]
\normalsize}
We finally combine the similarity values over multiple attributes by taking their mean value: 

\vspace{-3mm}
{\small
\[
sim(t_i, t_j) = \frac{\sum_{A \in \mathcal{M}_{attr}}sim(t_i.A, t_j.A) }{|\mathcal{M}_{attr}|}
\]
\normalsize}

\looseness-1
After computing the pair-wise similarity for tuples in the canonical relations,
we generate the initial tuple matches and their probabilities with the above 
similarity-to-probability method. In particular, we divide the tuple matches
into $50$ buckets and we use the evidence mapping in the gold
standard to label a sample of matches and produce the probabilities of the buckets. The sizes of the initial
tuple matches for each of the datasets are shown in Figure~\ref{tab:datastats}.

\begin{figure*}[th]
\begin{minipage}{0.33\textwidth}
\centering
\begin{subfigure}[b]{\linewidth}
\includegraphics[width=0.99\textwidth]{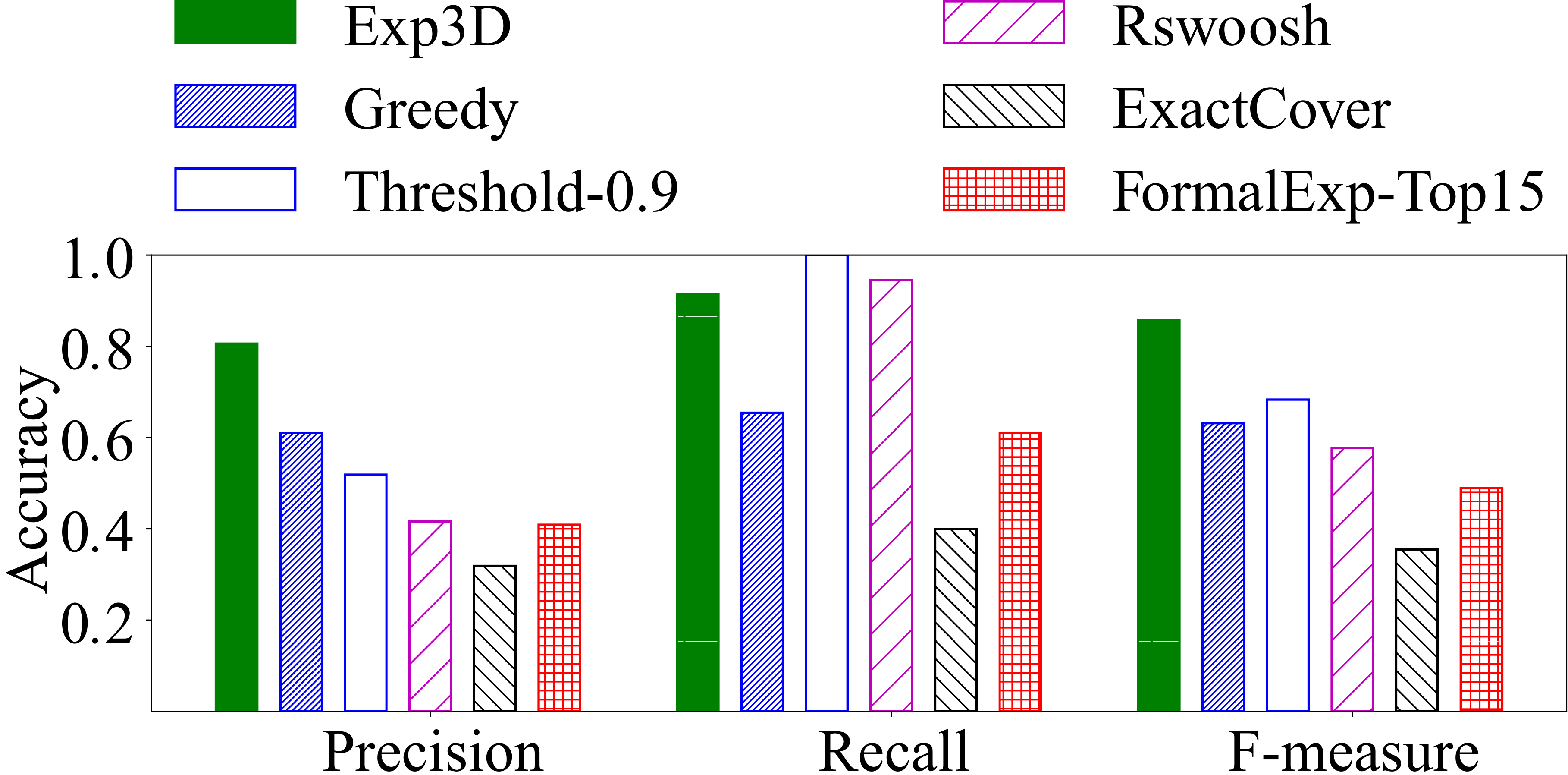}
\caption{NCES vs. UMass Explanation Accuracy.}
\label{fig:umass:exp}
\end{subfigure}
\end{minipage}
\begin{minipage}{0.33\textwidth}
\centering
\begin{subfigure}[b]{\linewidth}
\includegraphics[width=0.99\textwidth]{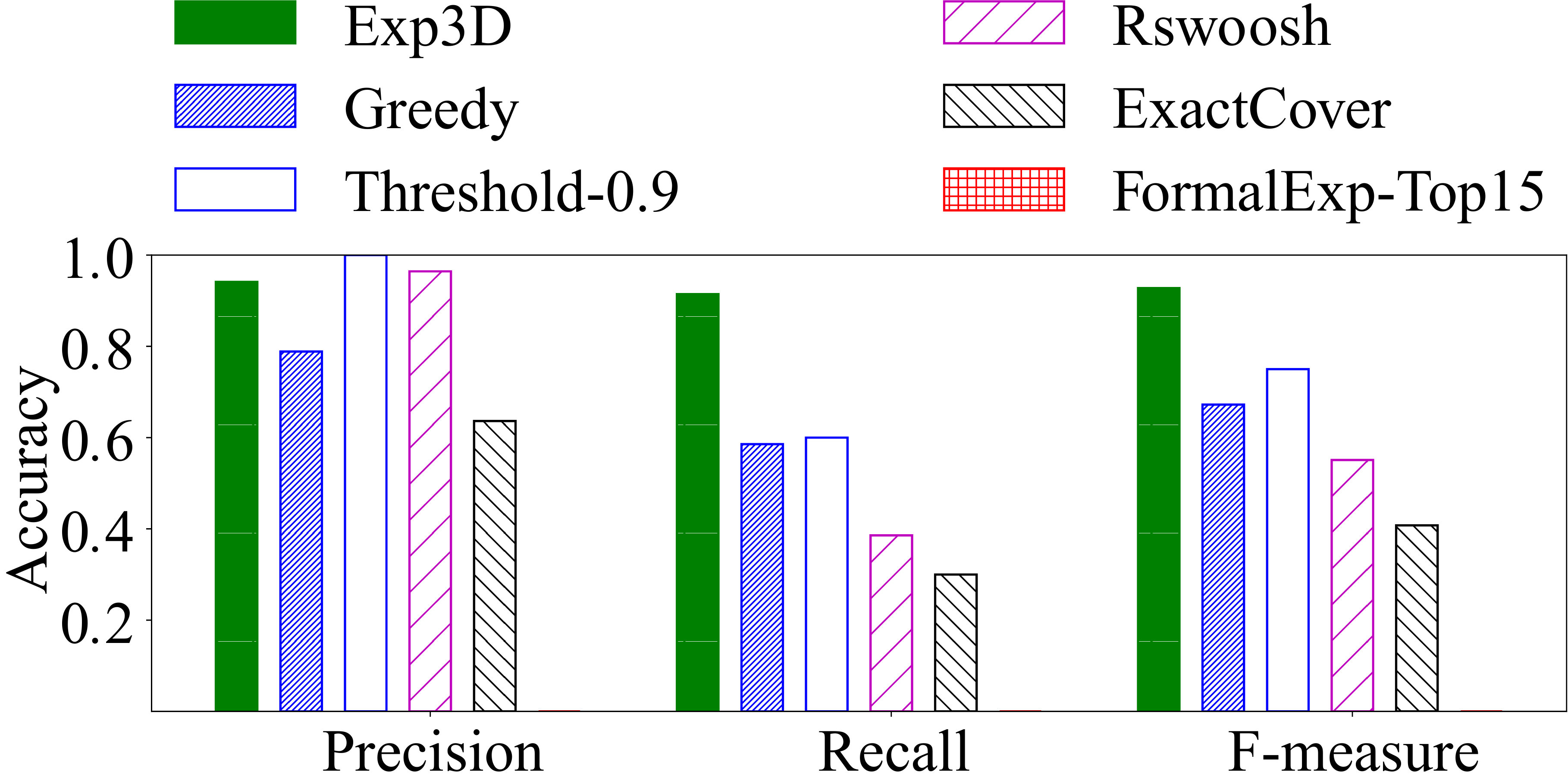}
\caption{NCES vs. UMass Evidence Accuracy.}
\label{fig:umass:edv}
\end{subfigure}
\end{minipage}
\begin{minipage}{0.33\textwidth}
\centering
\begin{subfigure}[b]{\linewidth}
        \hspace{1em}
\small
\begin{tabular}[b]{c|c} 
\toprule
\belowrulesepcolor{LightGray}
\rowcolor{LightGray}
\multicolumn{1}{c}{\textbf{Method}} & \textbf{NCES/UMass (sec)}  \\ 
\aboverulesepcolor{LightGray}
\midrule
\formal-Top15 & 0.052\\ \hline
\serf & 0.273 \\ \hline
\threshold-0.9 & 0.276 \\\hline
\greedy & 0.280\\ \hline
\exactcover & 0.272 \\\hline
\expdiff  & 0.322\\
\bottomrule 
\multicolumn{2}{l}{}
\end{tabular}
\caption{NCES vs. UMass Execution Time.}
\label{fig:umass:time}
\end{subfigure}
\end{minipage}\\
\begin{minipage}{0.33\textwidth}
\centering
\begin{subfigure}[b]{\linewidth}
\includegraphics[width=0.99\textwidth]{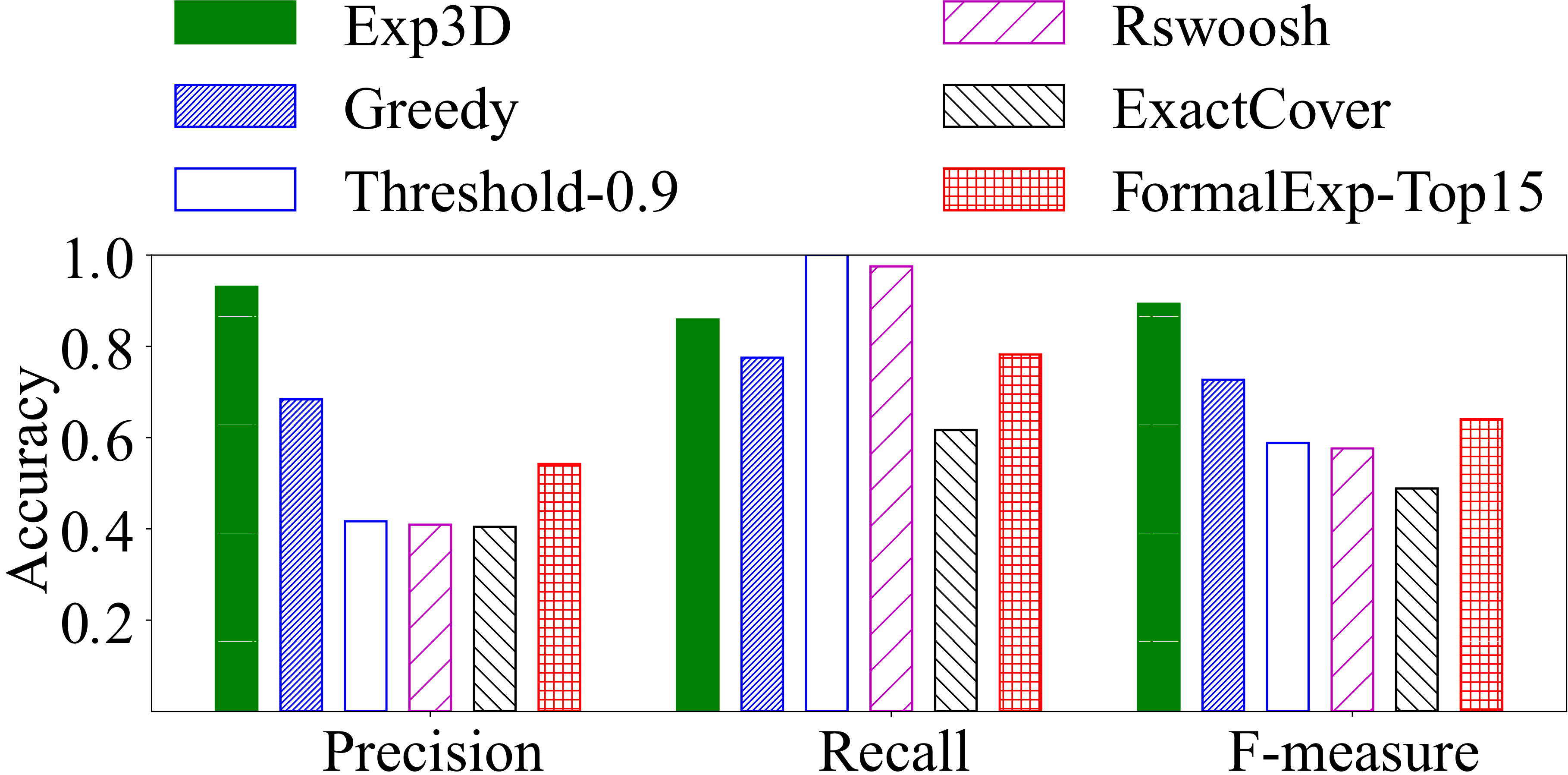}
\caption{NCES vs. OSU Explanation Accuracy.}
\label{fig:osu:exp}
\end{subfigure}
\end{minipage}
\begin{minipage}{0.33\textwidth}
\centering
\begin{subfigure}[b]{\linewidth}
\includegraphics[width=0.99\textwidth]{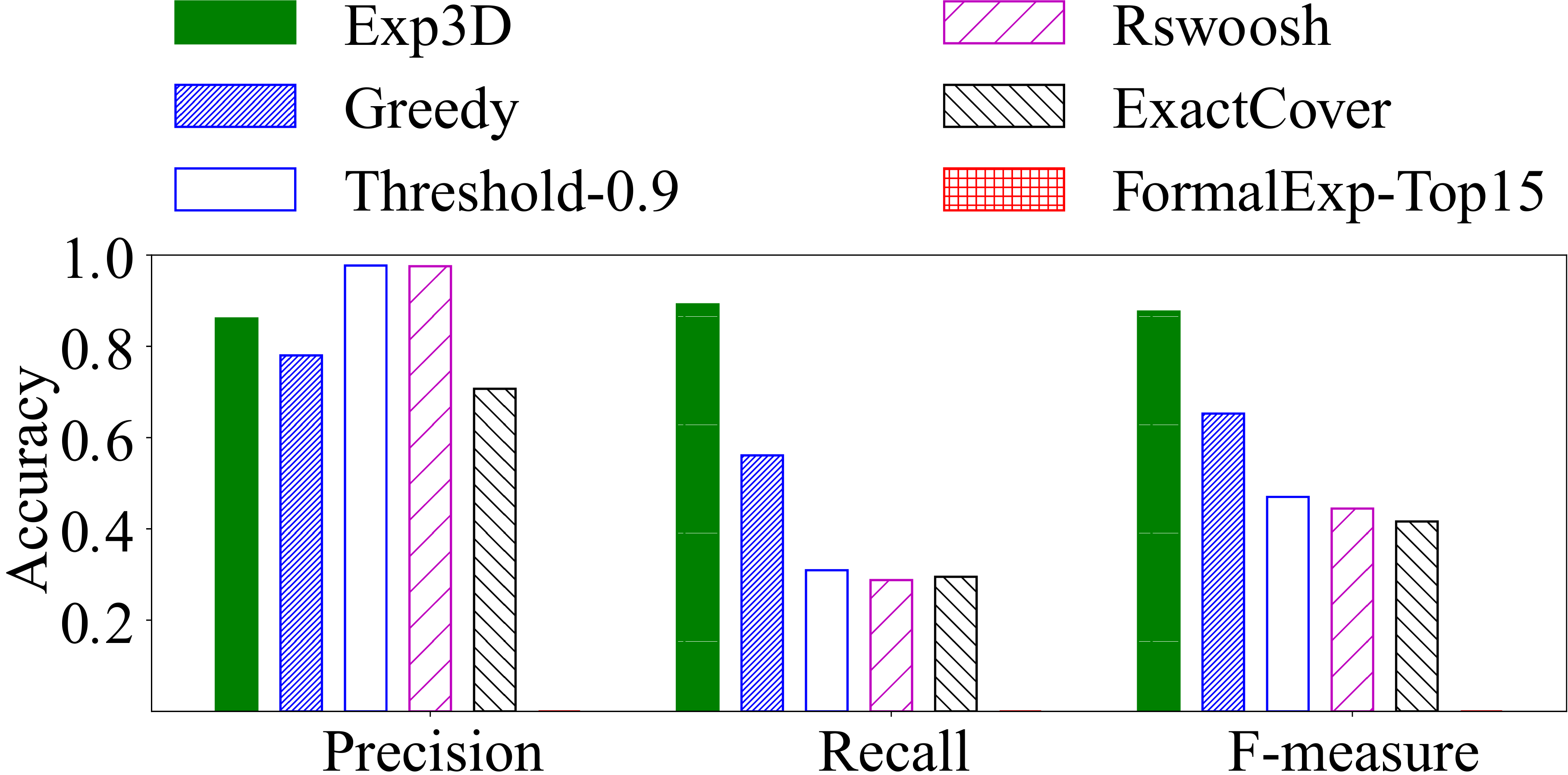}
\caption{NCES vs. OSU Evidence Accuracy.}
\label{fig:osu:edv}
\end{subfigure}
\end{minipage}
\begin{minipage}{0.33\textwidth}
\centering
\begin{subfigure}[b]{\linewidth}
    \hspace{1em}
\small
\begin{tabular}[b]{c|c} 
\toprule
\belowrulesepcolor{LightGray}
\rowcolor{LightGray}
\multicolumn{1}{c}{\textbf{Method}} & \textbf{NCES/OSU (sec)}  \\ 
\aboverulesepcolor{LightGray}
\midrule
\formal-Top15 & 0.064\\ \hline
\serf & 0.541 \\ \hline
\threshold-0.9 & 0.581\\\hline
\greedy & 0.573\\ \hline
\exactcover & 0.562 \\\hline
\expdiff & 0.729\\
\bottomrule 
\multicolumn{2}{l}{} \\
\end{tabular}
\caption{NCES vs. OSU Execution Time.}
\label{fig:osu:time}
\end{subfigure}
\end{minipage}
\caption{Accuracy and efficiency comparison over Academic datasets. \expdiff
achieves much higher accuracy than the other methods. \threshold obtains
high precision but low recall in the derived evidence. 
\formal does not provide any tuple matches in the evidence. }
\label{fig:edu}
\end{figure*}

\begin{figure*}[th]
\begin{minipage}{0.33\textwidth}
\centering
\begin{subfigure}[b]{\linewidth}
\includegraphics[width=0.99\textwidth]{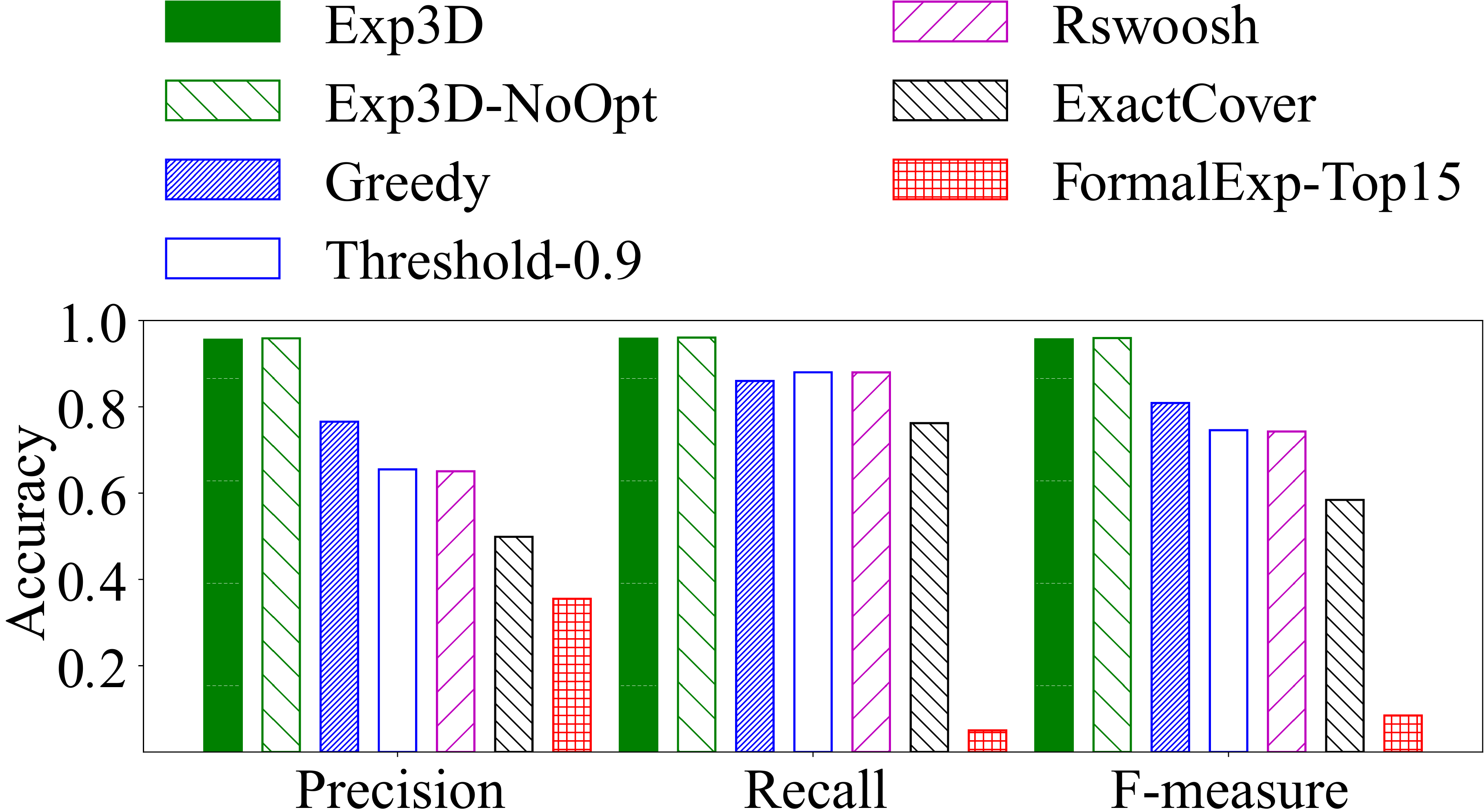}
\caption{ Average Explanation Accuracy.}
\label{fig:imdb:exp}
\end{subfigure}
\end{minipage}
\begin{minipage}{0.33\textwidth}
\centering
\begin{subfigure}[b]{\linewidth}
\includegraphics[width=0.99\textwidth]{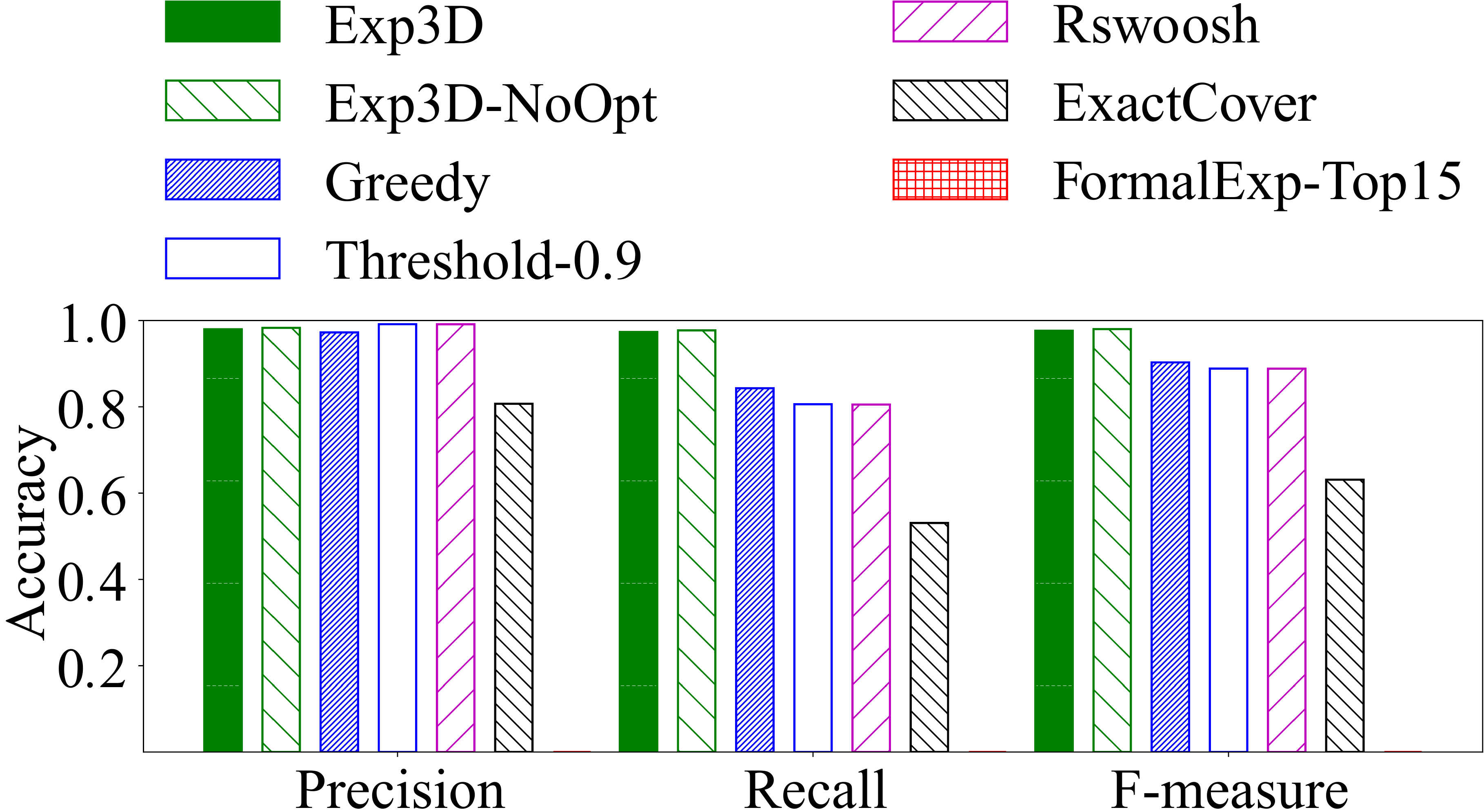}
\caption{Average  Evidence Accuracy.}
\label{fig:imdb:edv}
\end{subfigure}
\end{minipage}
\begin{minipage}{0.33\textwidth}
\centering
\begin{subfigure}[b]{\linewidth}
\includegraphics[width=0.99\textwidth]{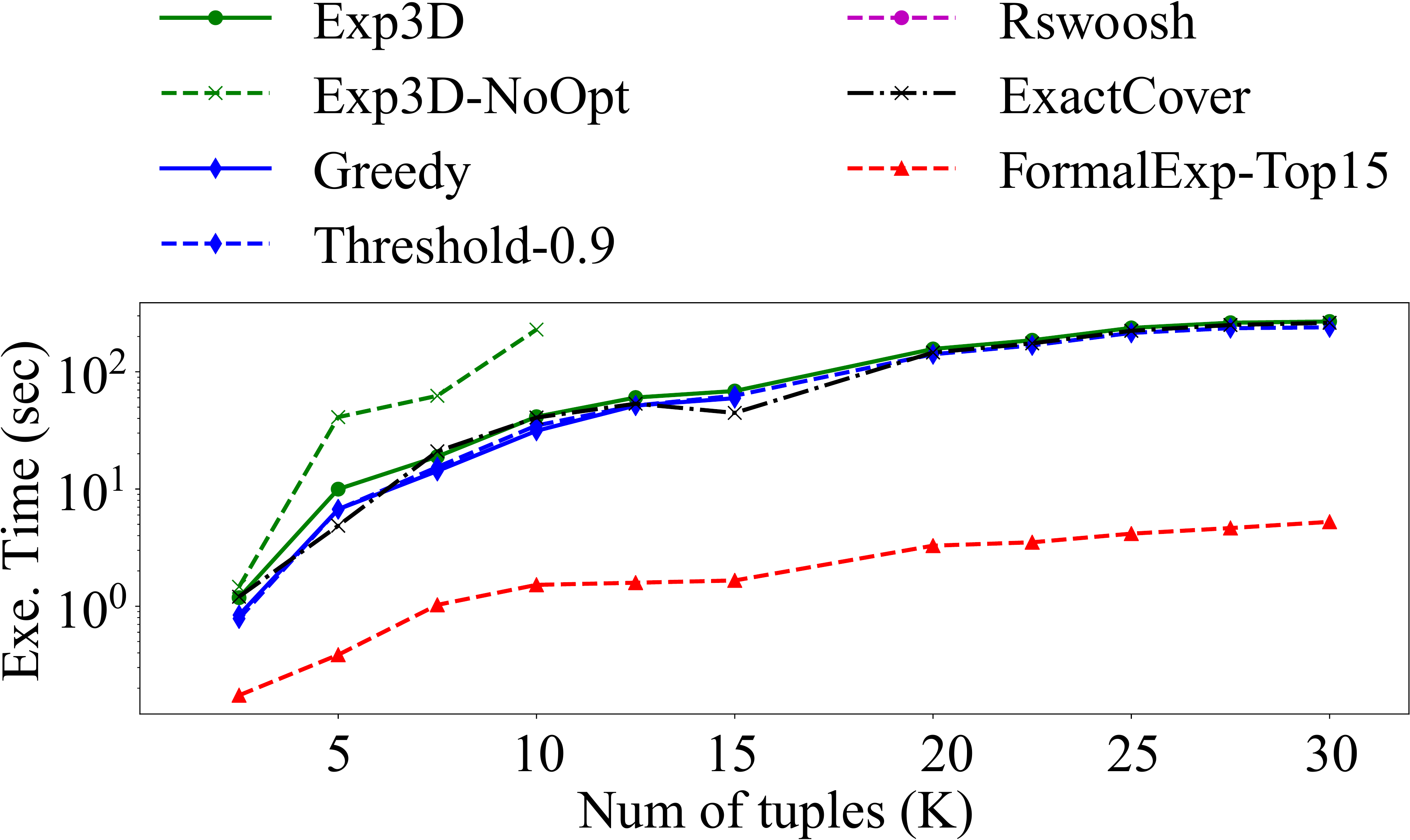}
\caption{Average Execution Time.}
\label{fig:imdb:time}
\end{subfigure}
\end{minipage}\\
\vspace{-2mm}
\caption{Accuracy and efficiency comparison over IMDb datasets. \expdiff achieves
near perfect accuracy. \serf and \expdiff without the smart-partitioning optimization fails 
to produce any results for queries with more than $10K$ tuples in 1hr. }
\label{fig:imdb}
\end{figure*}

\subsubsection{Algorithms}\label{sec:exp:methods}

We compare our framework, \system, against \formal, an approach that
focuses on explanations in the single dataset setting, \serf, a
state-of-the-art record linkage system, and three additional baseline methods. We
describe all the algorithms below.

\begin{description}[leftmargin=0mm, topsep=0mm, itemsep=0mm]
\item[\formal:] \formal explains surprising
outcomes of aggregate queries in a single database~\cite{roy2014formal}.
To apply \formal in disjoint datasets, we first compare the results
of the queries and then ask \formal to explain why the query result is high (or low)
on each individual dataset. 
Tuples that are included by the derived explanations are considered
provenance-based explanations. \formal returns the Top-$k$ explanations, and
requires $k$ as an input. In our experiments, we set $k=15$, denoted by
\formal-Top15, as it achieves the highest overall accuracy.

\item[\serf:] \serf~\cite{benjelloun2009swoosh} is an entity resolution technique that
produces deterministic tuple matches. For \serf, we treat all derived tuple matches
as the evidence mapping since their probabilities are all equal to $1.0$. We include tuples that do not have a match in this evidence mapping as provenance-based
explanations, and tuples with unequal impacts as value-based explanations. Here we use the Jaccard
similarity metric to compare string attributes, using $0.75$ as the default  threshold value.\footnote{We have also conducted
experiments using Jaro similarity, but its performance is strictly inferior to Jaccard similarity in
all experiments, so we don't report it in the graphs.}

\item[\threshold:] \threshold is a simple baseline that refines 
the initial probabilistic tuple matches by a fixed threshold.
It uses the derived evidence mapping to derive explanations, in the same manner as \serf.  In our experiment,
we set a threshold of $0.9$ and denote it as \threshold-0.9.

\looseness-1
\item[\greedy:] \greedy is a baseline that implements \system's objective
function (Definition~\ref{def:prob}), but builds the evidence mapping in a
greedy fashion (whereas \system derives it by solving constrained optimization
problems).
Initialized with an empty evidence mapping, \greedy
prioritizes tuple matches with higher probabilities and includes into
the evidence the match with highest probability that does not violate the valid mapping restriction 
(Definition~\ref{def:validmapping}) and improves the objective value. 
After examining all initial tuple matches, \greedy finalizes the evidence
mapping and creates the explanations in the same way as \serf and \threshold.

\looseness-1
\item[\exactcover:] 
We create a final baseline by adapting the integer programming solution of the
Exact Cover problem to solve the \probname problem as follows: we map tuples in one provenance
relation as elements, and tuples in the other provenance relation as sets; an element is covered by a set
if there exists an initial tuple mapping between their corresponding tuples. We further adapt the objective
function of the Exact Cover problem from a decision problem to a optimization problem, where
we want to find a collection of sets such that the total number of covered sets and elements
is maximized.

\looseness -1
\item[\expdiff:] Our proposed system, \system, expresses and optimizes the problem
as linear constraints and solves the constructed MILP problem(s) through a 
MILP solver (Section~\ref{sec:sol}, Section~\ref{sec:opt}). 

\end{description}
\subsubsection{Metrics}

\begin{description}[leftmargin=2mm, topsep=0mm, itemsep=0mm]
\item[Explanation accuracy:] We evaluate the explanation
accuracy of the algorithms using precision, recall, and
F-measure. We calculate precision as the fraction
of true explanations over derived explanations, and recall
as the fraction of true explanations over the gold standard;
F-measure is the harmonic mean 
of precision and recall ($\frac{2*precision*recall}{precision+recall}$).

\looseness-1
\item[Evidence accuracy:] We also evaluate the evidence mapping
accuracy with the same metrics. Similarly, 
we calculate the precision as the fraction of true tuple matches over
the refined tuple matches, and recall as the fraction of true matches
over the gold standard; F-measure as the harmonic mean of precision and recall.

\item[Execution time:] We evaluate the 
efficiency of all alternative algorithms through their 
total execution times, including the time for generating initial tuple matches.
\end{description}

\subsection{Real-world datasets} \label{exp:real}
We evaluate all the algorithms (Section~\ref{sec:exp:methods}) 
on both the Academic and IMDb datasets.
Figures~\ref{fig:umass:exp},~\ref{fig:osu:exp}, and~\ref{fig:imdb:exp}
demonstrate the precision, recall, and F-measure of the derived explanations; 
Figures~\ref{fig:umass:edv},~\ref{fig:osu:edv}, and~\ref{fig:imdb:edv}
demonstrate the precision, recall, and F-measure of derived evidence mapping; 
Figures~\ref{fig:umass:time},~\ref{fig:osu:time}, and~\ref{fig:imdb:time} demonstrate
the total execution time.

\smallskip

\noindent
\textbf{Single-dataset explanations.} 
Our evaluation with \formal examines whether single-dataset explanation
solutions could address explanations across different datasets. This method
does not generate an evidence mapping, and the derived explanations focus on
why a query result is high or low, rather than why it is higher or lower than
the other corresponding query. The why-high/why-low explanation question is
a best-effort adaptation of this solution to our problem setting, but it is not a
good enough proxy of the correspondence information encoded in the queries. As a result,
the f-measure of \formal-Top15 is low, indicating that it is ill-suited for
this problem setting.

\noindent
\textbf{Record-linkage approaches.} 
Record-linkage methods do not generate explanations as a goal, but the tuple
mappings they produce can be used as an evidence mapping and then mapped to
explanations. \serf and \threshold-0.9 produce evidence mappings with very
high precision because they employ thresholds in refining the mappings (thus
maintaining the most likely ones). However, their recall is low because they
eliminate correct mappings that happen to have low probabilities. As these
techniques miss many correct mappings, they include a large number of tuples
in the explanations, thus resulting in low explanation precision.  

Since \serf and \threshold-0.9 employ thresholds in refining the mappings,
they perform better when the initial mappings are of better quality, as is the
case for the IMDb datasets. Their performance drops significantly in the
Academic datasets. Through manual analysis, we noted that the initial tuple
mappings in the academic data misses or has low probabilities for a
significant portion of true matches. For example, the true tuple mapping,
(``Foodservice Systems Administration'', ``Food Business Management'') is
absent from the initial mapping. Such cases are common in the academic
datasets, but uncommon in the IMDb data because movie titles, persons' names,
and other attributes are less ambiguous. Further, our view generation and
error injection only contributed relatively small perturbations, making
matches easier to identify with higher accuracy.

\greedy is also a record linkage approach, but uses our objective function
instead of a strict threshold; thus, it is able to identify a larger portion
of true mappings and has a higher recall. However, it may easily reach a local
maximum, which results in lower precision and recall on the evidence mapping
and further hurts the explanation accuracy. \greedy is also impacted by the
initial mapping quality, but is a bit more robust to it compared to \serf.

Ultimately, record linkage methods also are oblivious to the correspondence
implied by the input queries. Failing to leverage this information, their
effectiveness remains relatively low (below 0.8 f-measure), even in the most
favorable data settings.

\noindent
\textbf{\expdiff.}
Our experiments demonstrate that our framework is highly accurate, with
respect to both explanations and evidence mappings. Its superior performance
compared to the other two categories of approaches is due to two main reasons.
First, its objective function is cognizant of the query associations, in that
it does not only focus on maximizing the quality of the matched tuples, but
also seeks to minimize the unmatched tuples. As a result, it produces smaller
explanations and identifies more correct mappings. As an example of the
distinction from record linkage, consider two datasets of two tuples: $A$, $B$
and $A'$, $B'$. Suppose that the initial probabilistic tuple mapping is $\{(A,
A', 0.8)$, $(B, B', 0.8)$, $(A, B', 0.9)$, and $(B, A', 0.5)\}$. Typical
record linkage methods would select $(A, B')$ as the single match, because it
maximizes the probability of the matched tuples. In contrast, \expdiff will
derive the correct true mappings, $(A, A')$ and $(B, B')$, because it
considers explanation optimality by avoiding un-matched tuples.
Second, record linkage methods often consider unmatched values as a very
negative signal for matching a pair of tuples. In contrast, \expdiff does not
weigh these mismatches as negatively, as it considers them as possible
value-based explanations. As a result, \expdiff is more robust to variations
in the quality of the initial tuple mapping. Nevertheless, the quality of the
initial mapping does play a role, thus \expdiff performs better on the IMDb
data than the academic datasets. However, in all cases, its accuracy is
superior to the other methods.

\looseness-1
Finally, while Exact Cover relates to \probname through the NP-completeness
reduction, it performs badly in all settings. This is expected since the
Exact Cover problem does not consider tuple impacts, and does not refine the
quality of the initial tuple mappings.

\smallskip

\looseness -1
\noindent \textbf{Efficiency.~} We show the total execution time of all methods in 
Figures~\ref{fig:umass:time},~\ref{fig:osu:time}, and~\ref{fig:imdb:time}.
All methods are very efficient, with under a second runtimes.
\threshold, \greedy,  \serf, \exactcover, and \expdiff rely on the same procedure
to derive the input tuple matches, which takes more than $98\%$ of their
total execution time. \exactcover scales better than the unoptimized version of  \expdiff, because
it has simpler problem settings.
Figure~\ref{fig:imdb:time} also demonstrates the effect of partitioning on the
IMDb data. Partitioning allows \expdiff to scale effectively, without impact
on its accuracy (Figures~\ref{fig:imdb:exp} and~\ref{fig:imdb:edv}).

\begin{figure*}[th]
\begin{minipage}{0.33\textwidth}
\centering
\begin{subfigure}[b]{\linewidth}
\includegraphics[width=0.85\textwidth]{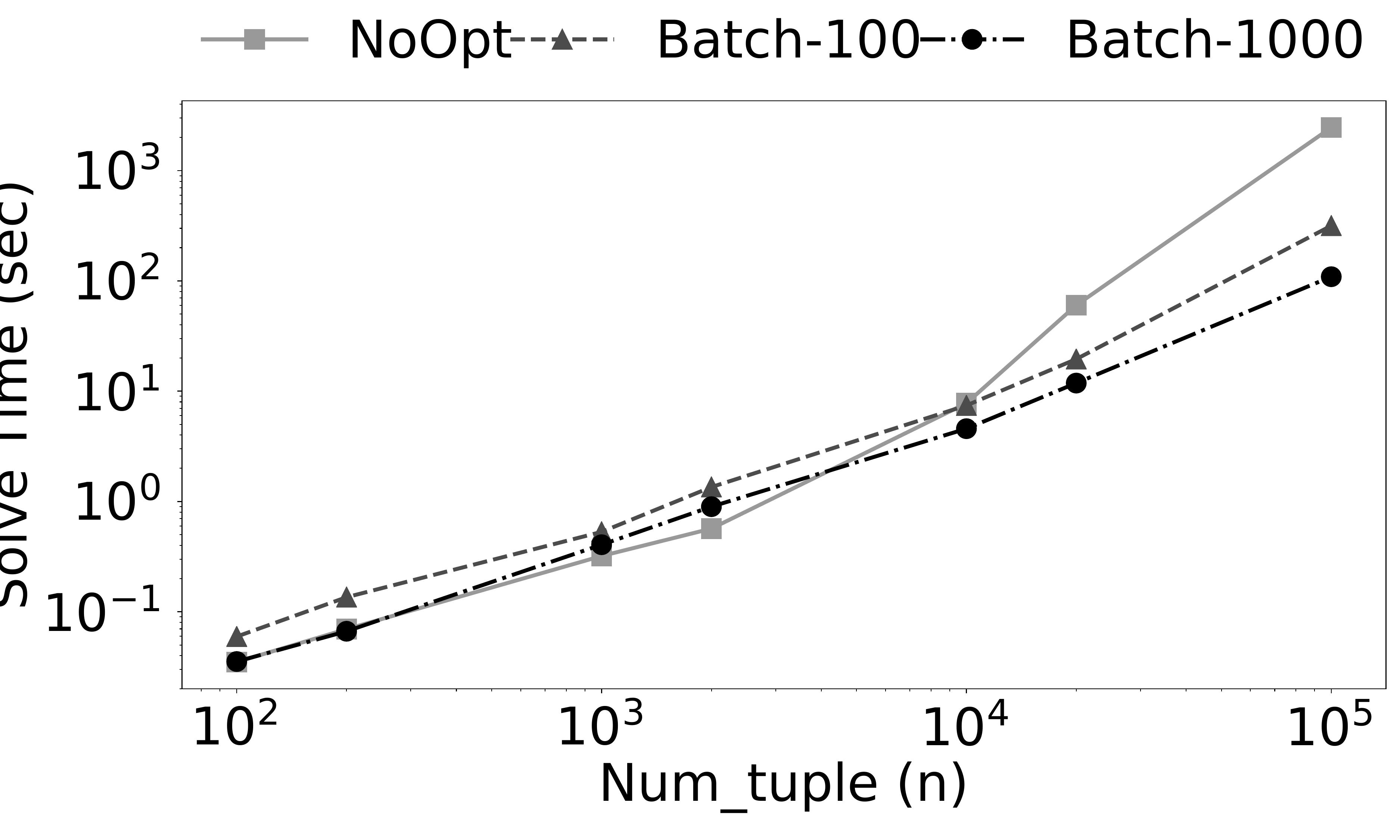}
\vspace{-2mm}
\caption{Solve time vs. \# of tuples (n).}
\label{fig:synthetic:n}
\end{subfigure}
\end{minipage}
\begin{minipage}{0.33\textwidth}
\centering
\begin{subfigure}[b]{\linewidth}
\includegraphics[width=0.85\textwidth]{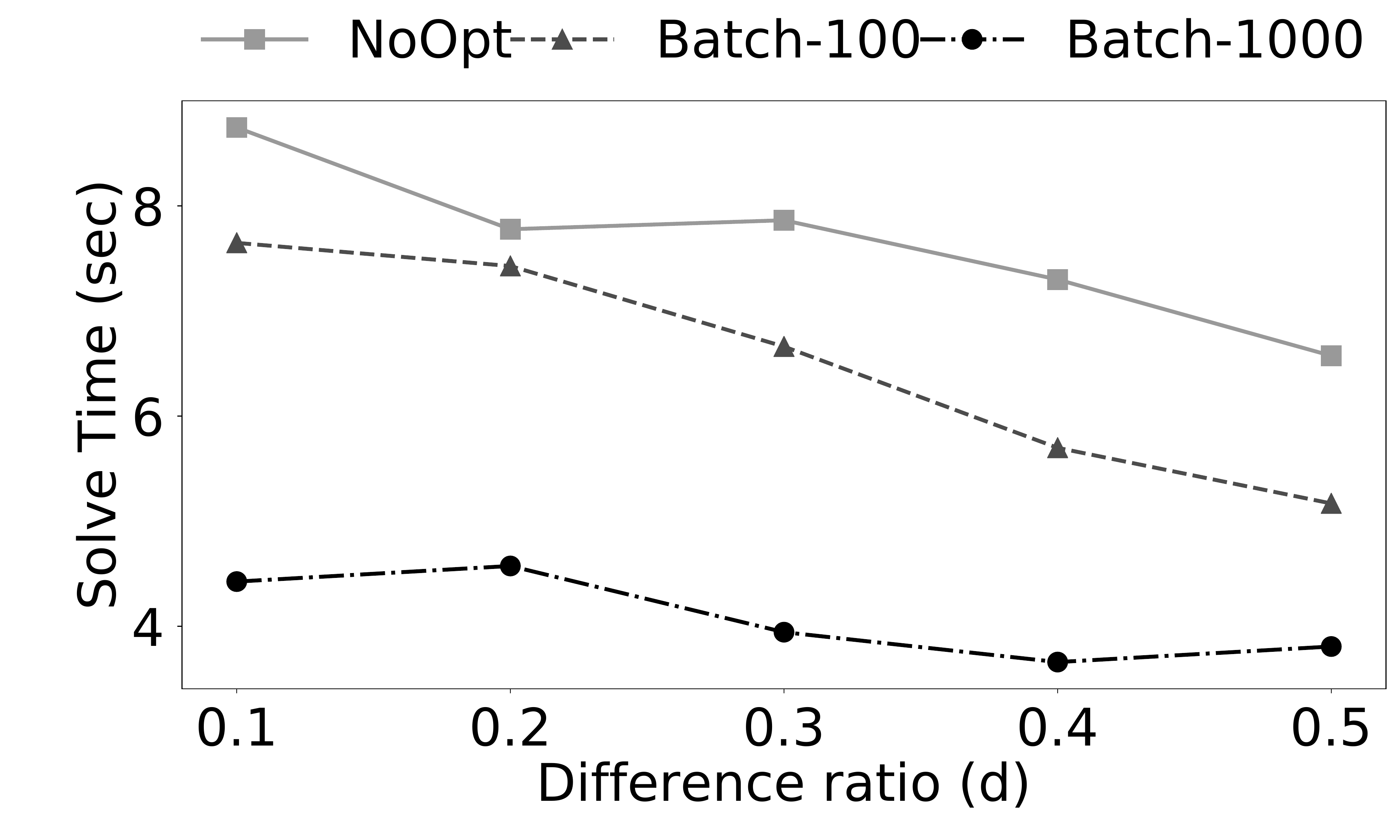}
\vspace{-2mm}
\caption{Solve time vs. difference ratio (p).}
\label{fig:synthetic:p}
\end{subfigure}
\end{minipage}
\begin{minipage}{0.33\textwidth}
\centering
\begin{subfigure}[b]{\linewidth}
\includegraphics[width=0.85\textwidth]{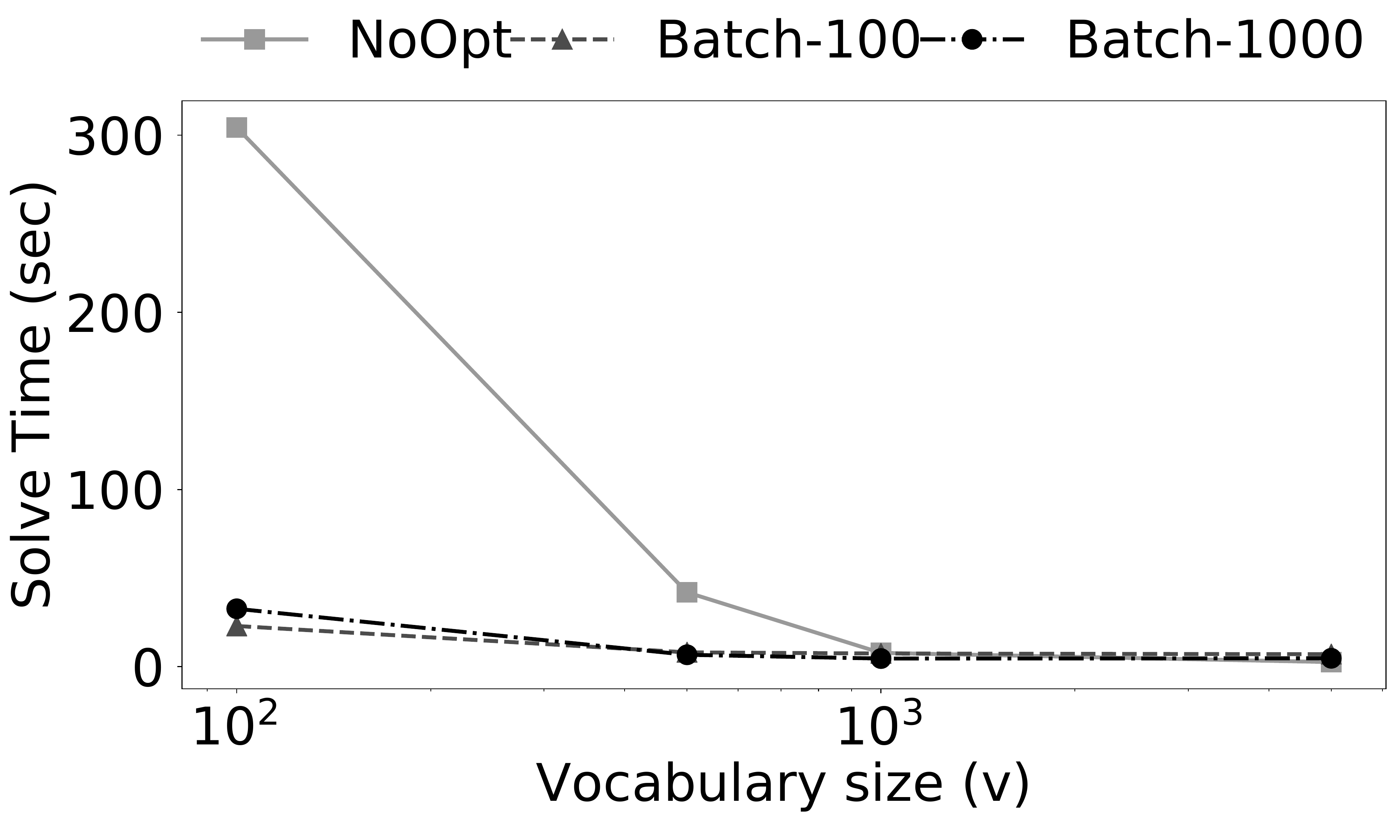}
\vspace{-2mm}
\caption{Solve time vs. vocabulary size (k).}
\label{fig:synthetic:k}
\end{subfigure}
\end{minipage}
\caption{Efficiency performance of \noopt, \bl, and \bh over synthetic datasets 
with diverse properties. Note that we only evaluate the solve time instead of the total execution time since the all methods share the same initial tuple matches generation time.}
\label{fig:synthetic}
\vspace{-2mm}
\end{figure*}

\subsection{Synthetic datasets} \label{exp:synthetic}
To stress-test \expdiff and evaluate its
smart-partitioning optimization, we create a synthetic data generator to produce
datasets and queries with diverse properties. In the synthetic data generator, we use
the same schema and queries for every pair of datasets:

\smallskip
\noindent
{\small \it
\resizebox{\columnwidth}{!}{
\begin{tabular}{ l l}
\toprule
\textbf{Dataset 1} &
\textbf{Dataset 2}\\
\midrule
Table(\underline{id}, match\_attr, val)  & Table(\underline{id}, match\_attr, val) \\
\multicolumn{2}{c}{ (match\_attr) $\equiv$ (match\_attr)}\\
$Q_{1}:$ & $Q_{2}:$\\
\texttt{SELECT SUM(val) FROM Table; } & \texttt{SELECT SUM(val) FROM Table; } \\\hline
\end{tabular}}}

\smallskip

\looseness -1
Based on the above schema, we follow three steps to 
produce a pair of datasets with the specified properties:
(1)~We first create $n$ tuples 
with random attribute values and add them to both datasets. (2)~We then
randomly drop $d$ percent of tuples, with uniform probability across tuples. 
(3)~We randomly select $d$ percent of tuples, again with uniform probability, 
and corrupt the tuples' ``val'' attribute. To generate random values in
the ``match\_attr'' attribute, we first create a vocabulary containing $v > 5$ random words and 
then generate phrases, each of which consists of $5$ random words from the vocabulary, as the attribute values; 
To generate random values in 
the ``val'' attribute, we randomly select an integer in the range of $[1, 10]$. 
The optimal explanations include tuples we dropped
or corrupted in the steps~(2) and~(3); the optimal evidence can be easily 
derived from step~(1).
In this experiment, we study the performance of smart-partitioning by dynamically changing the number of partitions ($k\in \mathbb{N}_{>1}$, Definition~\ref{prob:partition}) using a fixed batch size: $k = \lceil \frac{|T_1| + |T_2|}{batch\_size} \rceil$.

We evaluate \expdiff on three different settings: (1)~the basic algorithm without the smart-partitioning optimization (\noopt),
(2)~the optimized algorithm with batch size 100 (\bl), and (3)~the optimized algorithm with batch size 1000 (\bh). Figure~\ref{fig:synthetic} 
demonstrates the performance of \noopt, \bl, and \bh over diverse parameter settings.

\begin{description}[leftmargin=0mm, topsep=0mm, itemsep=-1mm]
\looseness -1
\item[Adjusting number of tuples ($n$):] We first adjust the number of tuples ($n$) in the synthetic 
datasets from $100$ to $100K$ with fixed difference ratio $d = 0.2$ and vocabulary size $v = 1K$. 
As shown in Figure~\ref{fig:synthetic:n}, \noopt performs well for problems with fewer 
tuples as the problem can be efficiently solved by a single MILP problem. However, its execution
time grows quadratically, if not exponentially, with increasing data size. 
\bl and \bh solve multiple MILP problems with bounded sizes, thus their solve time
grows linearly with increasing number of tuples. Meanwhile,
\bh is significantly more efficient than \bl as \bl requires longer time to
initialize and solve each individual sub-problems. With the smart-partitioning
optimization, \bh is more than $20\times$ faster than \noopt on problems with $100K$ tuples.

\item[Adjusting difference ratio ($d$):] We next adjust the difference ratio ($d$) from
$0.1$ to $0.5$ while keeping the other parameters fixed: $n = 1K, v = 1K$.
As expected, all three methods require longer time for problems with lower difference
ratio. This is because with higher difference ratio, there will be fewer tuples remaining
in the datasets. Again, \bh is much more efficient than \bl and \noopt.

\item[Adjusting vocabulary size ($v$):] Finally, we adjust the vocabulary size ($v$) from $100$
to $10K$ and keep $n = 1K, d = 0.2$. In the synthetic data generator, we generate
the value of the ``match\_attr'' attribute by randomly selecting $5$ words from the vocabulary. 
Thus the probability that two tuples share at least one common word increases
with lower vocabulary sizes. In other words, there will be many more initial tuple
matches when we set $v=100$ than $v = 10K$. 
As shown in Figure~\ref{fig:synthetic:k}, \bl is $15\times$ faster than \noopt and even outperforms \bh when $v=100$. This is because the number of tuple matches in each sub-problem also affects
the problem's overall complexity. Thus, we need to divide the problem into smaller partitions
when there is a larger number of initial tuple matches.
With increasing vocabulary size (and decreasing number of tuple matches), 
\bh starts to outperform the other two methods. When we increase the 
vocabulary size to a large enough number, e.g., $v = 10K$, \noopt, \bh, \bl start to perform similarly. 
\end{description}

In all experiments on the synthetic datasets,  \noopt, \bl, and \bh achieve near perfect accuracy 
in the derived explanations and evidence mapping.

\section{Related Work}
\label{s:related}
In this paper, we study the problem of explaining the disagreements in the
results of semantically similar queries over disjoint datasets. While there is
a growing body of work in data management research on deriving explanations,
existing work focuses on one dataset at a time, and cannot address
disagreements across datasets with potentially different schemas. \System is,
to the best of our knowledge, the first framework of its kind, that handles
disagreements across disjoint datasets.

Data management research on explanations has
focused on the assumption that data resides in a single dataset.
The Scorpion
system~\cite{Wu13} finds predicates on the input data as explanations for a
labeled set of outlier points in an aggregate query over a single
relation. Roy and Suciu~\cite{roy2014formal} extended explanations with a formal
framework that handles complex SQL queries and database schemas involving
multiple relations and functional dependencies. This explanation tool
does not require any preparation for the data and derives
the explanations as a set of conjunctive predicates.
Roy, Orr and Suciu~\cite{Roy2015} further extend their work to provide 
richer and more insightful explanations on datasets with prepared
candidate explanations derived by domain experts. 

Other explanation work investigates the absence of answers from a query
result~\cite{Chapman2009, tran2010conquer, tenCate2015}; these
systems provide why-not explanations and sometimes modification suggestions to
the queries. 
Work on provenance and causality~\cite{MeliouGMS2011, FreireGIM15,
Buneman2001} focuses on identifying the tuples that contribute to a query, and
quantify their contributions.
Finally, application-specific explanations focus on a particular
domain, such as performance of MapReduce jobs~\cite{Khoussainova2012}, item
rating~\cite{Thirumuruganathan2012, Das2011}, and auditing and
security~\cite{Fabbri2011, Bender14}.

To compare two semantically similar queries and the corresponding
databases, \system leverages existing schema matching techniques~\cite{hansen2000global, berlin2002database, madhavan2001generic, dhamankar2004imap}
to derive the correspondence among attributes in two semantically correlated schemas.
Existing schema matching solutions leverage a wide variety of 
techniques, from heuristics~\cite{dhamankar2004imap}, to rules~\cite{madhavan2001generic}, to learning-based approaches~\cite{hansen2000global, berlin2002database}. 

Another essential input for \system is the initial tuple matches (or the tuple mapping).
We may acquire such initial tuple matches by leveraging existing
entity resolution (or record linkage) techniques~\cite{davis2005establishing, benjelloun2009swoosh,
whang2010entity, bhattacharya2006latent,domingos2004multi}.
More specifically, \system treats existing entity 
resolution approaches
as blackboxes and uses them to derive the matches and include them
as part of the input.

\section{Summary of contributions}
\label{sec:conclusion}

In this paper, we presented an effective and scalable framework, \system, that
derives explanations for the disagreements between the results of two
semantically similar queries over two disjoint datasets. Our work formalized
several important concepts and essential properties that explanations should
satisfy.
\System uses a novel formalization and models explanations as two generic
types, provenance-based explanations and value-based explanations, and
evaluates the quality of explanations through a probabilistic model. The core
stage of \system is a translation of the explanation problem into a mixed
integer linear program, allowing the use of modern constrained solvers to
address it. Our work further introduced a smart-partitioning optimization that
allows \system to scale to large data sizes.
To the best of our knowledge, \system is the first explanation framework that can address disagreeing query results across disjoint datasets. 

\smallskip

\noindent \textbf{Acknowledgements:} This material is based upon work supported by the NSF under grants CCF-1763423 and IIS-1453543.

{
\balance
\bibliographystyle{abbrv}
\bibliography{expdiff}  

}

\end{document}